\documentclass{aa}
\usepackage{amssymb}
\usepackage{amsmath}
\usepackage{txfonts}
\usepackage{graphicx}
\usepackage{natbib}
\usepackage{rotating}

\bibpunct{(}{)}{;}{a}{}{,} 

\begin{document}

\title{Long Term Variability of
  Cyg~X-1\thanks{Table~\ref{tab:fitresults} is available in electronic
  form via http://www.edpsciences.org.}}
\subtitle{I. X-ray spectral-temporal correlations in the hard state}

\author{K.~Pottschmidt\inst{1}\fnmsep%
\thanks{\emph{Present address:} Max-Planck-Institut f\"ur
  extraterrestrische Physik, Giessenbachstra\ss{}e 1, 85748 Garching,
  Germany, and  INTEGRAL Science Data Centre, Ch.\ d'\'Ecogia 16, 1290
  Versoix, Switzerland}
 \and J.~Wilms\inst{1} \and
  M.A.~Nowak\inst{2} \and G.G.~Pooley\inst{3}\and T.~Gleissner\inst{1} \and
  W.A.~Heindl\inst{4} \and D.M.~Smith\inst{5} 
\and R.~Remillard\inst{2}
\and R.~Staubert\inst{1}}  
\institute{Institut f\"ur Astronomie und Astrophysik -- Astronomie,
  University of T\"ubingen, Sand~1, 72076 T\"ubingen, 
  Germany 
\and
  Center for Space Research, Massachusetts Institute of Technology, 77
  Massachusetts Ave., Cambridge, MA 02139, U.S.A. 
\and
  Astrophysics Group, Cavendish Laboratory, Madingley Road,
  Cambridge CB3 OHE, England 
\and
  Center for Astronomy and Space Sciences, Code 0424, University of
  California at San Diego, La Jolla, CA 92093, U.S.A. 
\and
  Space Sciences Laboratory, University of California at Berkeley,
  Berkeley, CA 94720, U.S.A.} 

\offprints{Katja Pottschmidt,
\email{katja.pottschmidt@obs.unige.ch}}
\date{Received 30 January 2002 / Accepted 12 June 2003 } 

\abstract{We present the long term evolution of the timing properties of
  the black hole candidate \mbox{\object{Cygnus X-1}} in the
  0.002--128\,Hz frequency range as monitored from 1998 to 2001 with
  the Rossi X-ray Timing Explorer (RXTE). For most of this period the
  source was in its hard state. The power spectral density (PSD) is
  well modeled as the sum of four Lorentzians, which describe distinct
  broad noise components.  Before 1998 July, Cyg~X-1 was in a
  ``quiet'' hard state characterized primarily by the first three of
  these broad Lorentzians being dominant.  Around 1998 May, this
  behavior changed: the total fractional rms amplitude decreased, the
  peak frequencies of the Lorentzians increased, the average time lag
  slightly increased, and the X-ray spectrum softened.  The change in
  the timing parameters is mainly due to a strong decrease in the
  amplitude of the third Lorentzian. Since this event, an unusually
  large number of X-ray flares have been observed, which we classify
  as ``failed state transitions''. During these failed state
  transitions, the X-ray power spectrum changes to that of the
  intermediate state. Modeling this PSD with the four Lorentzians, we
  find that the first Lorentzian component is suppressed relative to
  the second and third Lorentzian during the state transitions. We
  also confirm our previous conclusion that the frequency-dependent
  time lags increase significantly in the 3.2--10\,Hz band during
  these transitions.  We confirm the interpretation of the flares as
  failed state transitions with observations from the 2001~January and
  2001~October soft states.  Both the behavior of the PSD and the
  X-ray lag suggest that some or all of the Lorentzian components are
  associated with the accretion disk corona responsible for the hard
  state spectrum.  We discuss the physical interpretation of our
  results.   }

\titlerunning{Long Term Variability of Cygnus X-1}
\authorrunning{K.~Pottschmidt et al.}

\maketitle

\section{Introduction}\label{sec:intro}

The existence of distinct states for the X-ray emission of black hole
binaries (BHCs) has been extensively documented for many systems
\citep[see, e.g.,][]{zhang:97a,wilms:01c,homan:01a}. A geometrically
thin, optically thick accretion disk is thought to be responsible for
the soft, less variable X-ray flux dominating the soft state. In the
hard state, an additional hot plasma, the ``accretion disk corona'',
is postulated to produce a hard, highly variable emission component
which is in part reprocessed by the cooler disk.  The hot plasma is
also believed to be the reservoir for an outflow which emits
synchrotron radiation at radio wavelengths (and possibly also in the
X-rays, see \citealt{markoff:03a} and references therein). While the
general observational properties of BHCs are now well established
\citep[see, e.g,][]{klis:95,fender:99c,remillard:01a}, their
unification in terms of a physical model is uncertain
\citep{zdziarski:99b,nowak:01b}.  Especially, there exists no
consistent model for explaining the complex short term variability on
time scales faster than 1000\,s, although several phenomenological models
have been discussed
\citep{boettcher:01a,kotov:01a,poutanen:01a,churazov:01a,nowak:01b,kazanas:97a}.
These models explain the evolution and propagation of fluctuations in
the X-ray flux in terms of several different mechanisms such as, e.g.,
infalling blobs of matter, development of localized X-ray flares,
reflection, or wave propagation.

One of the possibilities to further improve current accretion models
is to study the long term evolution of the canonical hard state, since
in this state all the important physical components -- disk, hot
plasma, synchrotron outflow -- as well as prominent short term
variability can be observed. The stability of its characteristic
parameters, however, has yet to be systematically studied with high
spectral and temporal resolution. We therefore have performed regular
monitoring observations of Cygnus~X-1 with RXTE from 1998 to 2001.
This persistent high mass X-ray binary (HMXB) is the best candidate
for such a hard state campaign: it is bright ($\sim$0.6\,Crab) and
spends most of its time in the hard state. In addition to the
canonical hard state properties, the 2--12\,keV X-ray flux and the
15\,GHz radio flux of Cyg~X-1 display the 5.6\,d orbital period and a
150\,d period which might be due to precession of a disk/jet system
\citep{pooley:98a,brocksopp:99a}. The hard state shows weak and
persistent radio emission that has been resolved as a type of steady
jet \citep{stirling:01a}. The radio emission turns off during the soft
state, as has been well documented during the last extended soft state
which occurred from 1996~May to 1996~August
\citep{zhang:97a,cui:97b,cui:97d,wen:01a}.

The hard state PSD of Cyg~X-1 has often been observed and modeled, the
source serving as the BHC prototype. In addition to the flat-topped
hard state noise, the occasional presence of low frequency noise on
timescales longer than 1000\,s and of a probably associated weak
quasi-periodic oscillation (QPO) at $\sim$$0.04$\,Hz has been reported
\citep{angelini:94,vikhlinin:94}. Early descriptions of the
flat-topped PSD attributed the variability to Poisson distributed
X-ray flares (shot noise) and generally approached the PSDs as being
due to a single broad-band process \citep[][see, however,
\citealt{belloni:90a}]{nolan:81,lochner:91,pottschmidt:98a}. In recent
years, however, it has become increasingly clear that the hard state
PSDs of Cyg~X-1 and other BHCs show several ``breaks'', i.e.,
frequencies where the PSD steepens \citep[][and references
therein]{nowak:98b,belloni:90b}, and that this complex structure has
to be described by multiple components
\citep{smith:97a,berger:98a,nowak:00a,nowak:01b,belloni:02a}. Often,
the basic component of such models is still a shot noise component
\citep{poutanen:01a,smith:97a}.

While on first sight the hard state PSD of Cyg X-1 is dominated by two
distinct broad components \citep{gilfanov:99a}, strong indications
were recently found that it can be described remarkably well with four
broad noise components that yield the main contribution to the root
mean square (rms) amplitude in the 0.001\,Hz to 100\,Hz range
\citep{nowak:00a}.  Occasionally, narrow features representing QPOs
are required.  A previously observed ``QPO'' peak, detected near 1\,Hz
with $Q\sim1$ and $\sim 10\%$ rms amplitude \citep{rutledge:99a}, can
possibly be identified with one of the above mentioned broad
components being very prominent. During transitions to and from the
soft state, broader QPOs ($Q\sim0.5$--1, 3--17\% rms amplitude) between
3 and 13\,Hz have been observed \citep{cui:97b}. As we find below,
their parameters suggest that they correspond to the noise components
in the hard state, with varying central frequencies and amplitudes.
The normalization and flat-top break frequency -- associated with the
broad component that has the lowest characteristic frequency -- have
long been known to be anti-correlated \citep{belloni:90a}.  In
addition, the two lowest characteristic frequencies associated with
broad components have been observed to be correlated
\citep{gilfanov:99a}.  This latter correlation was discovered by
\citet{wijnands:99a} by comparing the flat top break frequency with
the broad hard state ``QPOs''.

A four-peak deconvolution of hard state black hole PSDs is especially
interesting in the context of the question of whether any of these BHC
noise features have the same origin as the generally much narrower and
faster QPOs seen in neutron star binaries \citep[e.g., the so-called
horizontal branch QPO and twin kHz
QPOs;][]{wijnands:99a,psaltis:99a,nowak:00a,vanstraaten:01a}.
The aim of our PSD fits is therefore to identify the main noise
components and to characterize their evolution in order to provide new
data for constraining accretion models.

In this work we concentrate on describing the variability properties
of Cyg~X-1 in terms of the power spectral density (PSD), the
Fourier-frequency dependent X-ray time lags, and simplified spectral
models. Our aim is to characterize the accretion flow in the hard
state. The description of the X-ray spectra in terms of more
sophisticated accretion disk corona models will be published in a
separate paper \citep{gleiss:02a}.  This paper is organized as
follows: In Sect.~\ref{sec:data} we describe our observing campaign,
the data screening, the computation of the power spectra, and the
procedure of fitting multiple Lorentzian profiles to the power
spectra. We then present our results in Sect.~\ref{sec:evolution},
starting with the long term evolution of the power spectrum
(Sect.~\ref{sec:twohard}), followed by a discussion of the failed
state transitions (Sect.~\ref{sec:failed}), the transition into the
2002 soft state (Sect.~\ref{sec:02soft}), and the behavior of the
narrow Lorentzians (Sect.~\ref{sec:thin}).  In Sect.~\ref{sec:disc} we
summarize our results.

\section{Observations and Data Analysis}\label{sec:data}

\subsection{The RXTE Monitoring Observations of Cyg X-1}\label{subsec:monit}
The RXTE data presented here were obtained with the Proportional
Counter Array \citep[PCA;][]{jahoda:96} and with the All Sky Monitor
\citep[ASM;][]{levine:96}, using the pointed RXTE observation programs
P30157 (1998), P40099 (1999), and P50110 (2000--2001). We used the
standard RXTE data analysis software, \texttt{FTOOLS}, Version~5.0, to
create lightcurves with a time resolution of $2^{-6}$\,s (P30157) or
$2^{-8}$\,s (P40099, P50110) in multiple energy bands. The energy
bands extracted for P30157 and P40099/P50110 present the closest match
possible for the available data modes (see Table~\ref{tab:energies}),
and include the bands used in our analysis of the X-ray time lags
\citep{pottschmidt:00a,pottschmidt:00b}.  With a few exceptions noted
below, the PSD analyses presented here were performed using summed
lightcurves in the $\sim$2\,keV to 13.1\,keV energy range.

The P30157 data set (weekly monitoring in 1998 with a nominal exposure
time of 3\,ks per observation) is the same that was used previously for
the study of the evolution of X-ray time lags
\citep{pottschmidt:00a,pottschmidt:00b} and the evolution of X-ray
spectra \citep{smith:01b}. The data set P40099
(biweekly monitoring in 1999 with a nominal exposure time of 10\,ks
per observation) presented here has also been included in the time lag
study of \citet{pottschmidt:00b}. The P50110 data (biweekly
monitoring with a nominal exposure time of 20\,ks per observation)
have not been published before. Starting in 1999, simultaneous
observations at radio wavelengths performed with the Ryle telescope in
Cambridge, U.K., are also part of the campaign. In addition, daily
radio fluxes were obtained with the Ryle telescope.

For all observations, high resolution lightcurves and X-ray spectra were
generated using standard screening criteria: a pointing offset of $<0 \fdg
01$, a SAA exclusion time of 30 minutes\footnote{For a source as bright as
  Cyg~X-1 this is a very conservative criterion.}, and a source-elevation
of $>10\degr$. At times of low source luminosity, when the ``electron
ratio'', i.e., a measure for the particle background in the PCA, was not
influenced by the source itself, we also imposed a maximum ``electron
ratio'' of 0.1. Good-time intervals (GTIs) were then created for all
available PCU combinations, and from those suitable GTIs were selected for
extracting high time-resolution data for the data modes and energy bands
given in Table~\ref{tab:energies}.  For each GTI file, we generated the
background spectrum using the FTOOL \texttt{pcabackest} and the ``sky-VLE''
background model. The background in a given energy band was then taken into
account in determining the PSD normalization.

As a result of the data screening, $\sim$2\,ks of usable data were
left for each of the P30157 observations and $\sim$7\,ks for P40099
and P50110. Note that due to RXTE scheduling constraints the full
nominal exposure time was only rarely reached. Observation dates and
exposure times are included in Table~\ref{tab:fitresults}. The table
caption given here contains a description of all parameters, the full
table is available in electronic form at EDP Sciences.

\begin{table}
\caption{Energy bands and data modes used.}\label{tab:energies}  
\noindent\begin{tabular}{lrrr} 
            &          PSD   & \multicolumn{2}{r}{lag/coherence} \\ 
            &                & low          &  high  \\
\hline
\multicolumn{4}{l}{PCA Epoch~3: data taken until 1999 March 22}\\
\multicolumn{4}{l}{data mode B\_16ms\_46M\_0\_49\_H} \\
channels    &         0--35  &        0--10 &   23--35 \\ 
energy [keV]& $\sim$2--13.0  & $\sim$2--4.2 & 8.3--13.0\\ 
\hline
\multicolumn{4}{l}{PCA Epoch~4: data taken after 1999 March 22}\\
\multicolumn{4}{l}{data mode B\_2ms\_8B\_0\_35\_Q} \\ 
channels    &         0--30  &        0--10 &   20--30 \\ 
energy [keV]& $\sim$2--13.1  & $\sim$2--4.6 & 8.4--13.1\\ 
\hline
\end{tabular}

\end{table}

\subsection{Computation of the Power Spectra}\label{subsec:calc}
The computation of the power spectra for all energy bands given in
Table~\ref{tab:energies} follows \citet{nowak:98b} and
\citet{nowak:00a}. 

For the P30157 data set, PSDs in the 0.002 to 32\,Hz range were
computed, while the P40099 and P50110 PSDs reach up to 128\,Hz. This
discrepancy arises from the different maximum time resolutions of the
binned data modes that are available for these data sets (see
Table~\ref{tab:energies}). We decided to use the binned modes with
their comparatively moderate time resolutions, because they allow us to
study several different energy bands (see Sect.~\ref{sec:failed})
which cannot be done with the ``single bit'' data modes (250\,$\mu$s
time resolution) that are also available. 

All PSDs used in this paper are presented in the normalization of
\citet{belloni:90a} and \citet{miyamoto:92}, where the PSD integrated
over a certain frequency interval equals the square of the relative
contribution of that frequency interval to the total rms noise of the
source.  Where possible, we adjusted the normalization to the
background corrected count rate.  The observational deadtime corrected
Poisson noise was subtracted before the normalization. See
\citet{vikhlinin:94b} and \citet{zhangw:95a} for a discussion of the
deadtime influence on the PSD in general, \citet{jernigan:00a} and
\citet{zhangw:96a} for the case of the PCA detector, and
\citet{revnivtsev:00a} for an application to PCA measurements of
Cyg~X-1 above 100\,Hz.  For most observations we used the first two
components of Eq.~(6) of \citet{jernigan:00a}, namely the approximation
of the ``general paralyzable deadtime influence''
\citep[][Eq.~(24)]{zhangw:95a} plus the deadtime caused by ``Very Large
Events'' (characterized by the PCA VLE count rate).  For frequencies
$<$100\,Hz the correction of the noise level due to these two
components amount to only a few percent.  Higher order corrections or
a fit of these components to the PSDs as applied by
\citet{jernigan:00a} or \citet{revnivtsev:00a} are therefore not
necessary here. We note that the discussed features in the PSDs have
rms amplitudes well above the typical deadtime structures (see, e.g.,
Fig.~5 of \citealt{jernigan:00a} or Fig.~4 of
\citealt{revnivtsev:00a}).

A slightly different PSD normalization and deadtime correction has
been applied to all observations performed after 2000 May 12, when the
epoch of enhanced PCU~0 background count rates started. Here, a good
background model was not yet available and thus we could not
background correct the PSDs. This correction is only on the order of a
few percent and does not influence appreciably our results \citep[we
note that this problem is, however, a serious problem for sources with
a low count rate;][]{kalemci:01a}.  Furthermore, as we do not have
access to the VLE count rates from individual PCUs, it becomes
impossible to correct the PSD for the VLE specific deadtime.
Therefore, we only applied the general paralyzable deadtime
correction.

For each observation we created three PSDs by averaging the individual PSDs
from lightcurve segments with durations of 512\,s, 128\,s, and 32\,s.
Logarithmic frequency rebinning was performed on the same frequency grid
for each of the three resulting PSDs. This grid ranges from $2^{-9}$ to
$2^5$ (or $2^7$)\,Hz ($\sim$0.002 to 32 [or 128]\,Hz), with different
d$f/f$ values for different frequency ranges: 25 logarithmically spaced
bins were used from 0.001\,Hz--0.1\,Hz, 50 bins from 0.1--20\,Hz, and 5(15)
bins from 20--32(128)\,Hz. The final PSD was then created by selecting for
each grid bin the highest signal to noise value among the three PSDs.
  
\begin{figure}
\resizebox{\hsize}{!}{\includegraphics{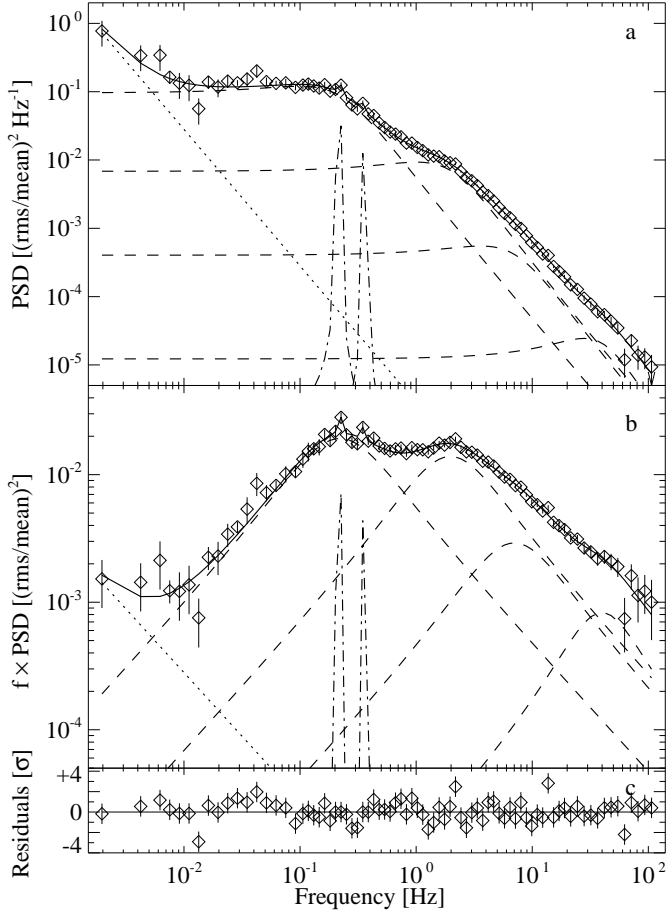}}
\caption{\textbf{a)} Best fit (solid line) to a typical 0.002--128\,Hz
  hard state power spectrum of Cyg X-1 (P40099/18, 14off, 1999
  September 13, diamonds) using multiple Lorentzians. The fit was
  performed for the energy range from $\sim$2 to 13.1\,keV. It
  includes four broad (dashed lines) plus two narrow (dash-dotted
  lines) Lorentzian profiles and a power law component (dotted line).
  \textbf{b)} The same as above but here the PSD has been multiplied by
  frequency to illustrate the peak frequencies of the broad noise
  components (see Sect.~\ref{sec:disc} for a discussion of the
  importance of these frequencies). \textbf{c)} Residuals between data
  and model in units of the standard deviation of the PSD data.}
\label{fig:fitex}
\end{figure}

\begin{figure}
\resizebox{\hsize}{!}{\includegraphics{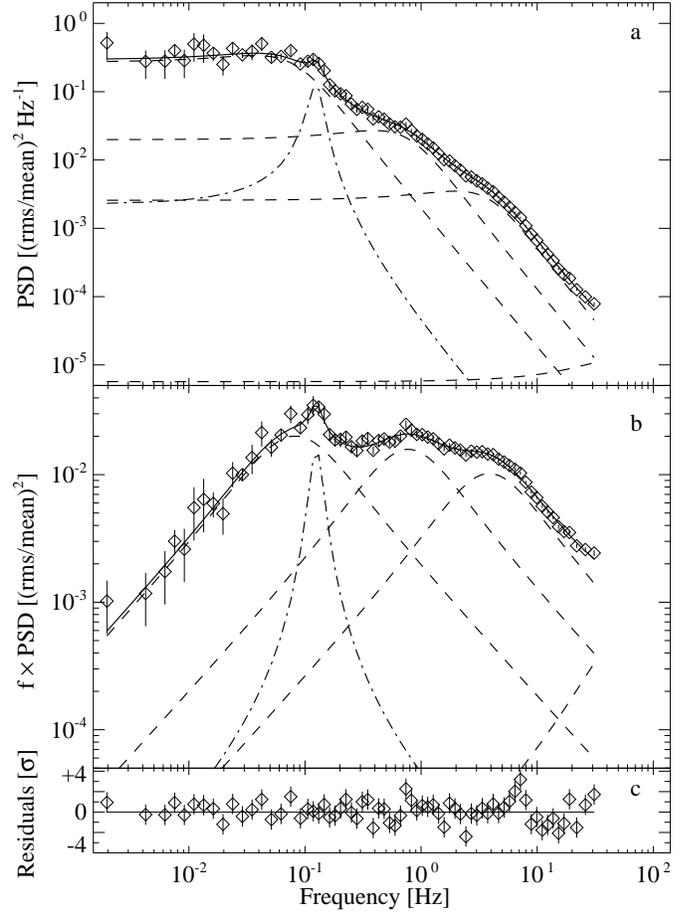}}
\caption{\textbf{a--c)} Another typical fit, in this case to one of the
  P30157 observations (No.\ 11, 1998 February 20) in the 0.002--32\,Hz
  frequency range. The best fit consists of four broad and one
  narrow Lorentzian profiles.}
\label{fig:preflare}
\end{figure}

\subsection{Modeling the Power Spectra}\label{fits}
As was shown by \citet{nowak:00a}, multiple Lorentzian profiles can
provide a good description of the typical hard state variability of
Cyg~X-1 and \object{GX~339$-$4} in the 0.001--100\,Hz range.  Inspired
by this success, we applied multiple Lorentzian profiles to model the
PSDs from our monitoring campaign and generally obtained good fits
(see Table~\ref{tab:fitresults} and the typical examples displayed in
Fig.~\ref{fig:fitex} and Fig.~\ref{fig:preflare}).  Note that power
spectra obtained during the soft state, e.g., during the 1996 soft
state, are not well fit with the multi-Lorentzian model (see also
Sect.~\ref{sec:failed} below).  Our fitting
approach uses standard $\chi^2$ minimization with the uncertainty of
the free fit parameters determined using the usual prescription of
\citet{lampton:76a}. Unless noted otherwise, we give the uncertainty
in terms of $1\sigma$ errors.

We describe the power spectra as the sum of Lorentzian profiles of the
form
\begin{equation}\label{eq:lorentzian}
L_i(f)=\pi^{-1}\frac{ 2 R_i^2 Q_i f_i}{{f_i}^2+4 Q_i^2(f-f_i)^2} 
\end{equation}
where $f_i$ is the resonance frequency of the Lorentzian, $Q_i\sim
f_i/\Delta f_{i, \text{FWHM}}$ its quality factor\footnote{$\Delta
  f_{i, \text{FWHM}}$ is the full frequency width of the Lorentzian at
  half of its maximum value.}, and $R_i$ its normalization constant.
Note that our definition of $Q$ is consistent with the common use of
the $Q$-value \citep[e.g.,][]{belloni:02a}, but differs from that used
by \citet{nowak:98b} and \citet{nowak:00a} where $Q=f_i/\Delta f_{i,
  \text{HWHM}}$ with $\Delta f_{i,\text{HWHM}} = \Delta
f_{i,\text{FWHM}}/2$ was used.

Integrating this profile over the frequency range from zero to
infinity gives its total rms amplitude:
\begin{equation}\label{eq:rms}
\text{rms}_i=R_i \left(\frac{1}{2}-\frac{\tan^{-1}(-2 Q_i)}{\pi}\right)^{1/2}
\end{equation}
A further important quantity of the Lorentzian is its peak frequency,
i.e., the frequency where its contribution to the total rms
variability is at its maximum:
\begin{equation}\label{eq:peak}
\nu_i= f_i \left( 1 + \frac{1}{4 Q_i^2}\right)^{1/2}
\end{equation}
As it was recently found that $\nu_i$ seems to be the important
parameter in terms of frequency correlations
\citep{nowak:00a,nowak:01b,vanstraaten:01a,belloni:02a}, we will use
$\nu_i$, and not $f_i$, to describe the location of each Lorentzian,
unless specifically noted otherwise.

\begin{table*}
\caption{Results of fitting multiple Lorentzians to selected power
  spectra of the Cyg~X-1 RXTE/PCA monitoring observations from
  1998--2001 in the frequency range from (1/512)\,Hz to 32\,Hz
  (P30157) or 128\,Hz (others), for the 2--13\,keV band
  (Table~\ref{tab:energies}). The table is available in electronic
  form via http://www.edpsciences.org. Apart from the fit 
  results, it contains the following information for each observation:
  running number, proposal name (e.g., ``P40099''), sub-ID within the
  proposal, PCUs turned off, date of the observation, and the exposure
  time in ks. The fit results for each observation are organized as
  follows: for each of the possible four broad and three thin
  Lorentzians the norm ($R$), central frequency ($f$), and quality
  factor ($Q$) with their corresponding uncertainties are listed. In
  addition the power law normalization ($A$) and slope ($\alpha$) are
  given if present in the model. Finally, the $\chi^2/{\text{dof}}$
  and $\chi^2_{\text{red}}$ values of the fit are
  listed.\label{tab:fitresults}} 
\end{table*}

Alternative models for the lowest frequency Lorentzian, such as using
a zero centered Lorentzian, i.e., a shot noise component featuring a
single relaxation time scale \citep{belloni:02a}, may in principle
provide fits of comparable quality \citep[note that in the case of the
observation analyzed by][a zero centered Lorentzian provided a
significantly worse fit]{nowak:00a}. The multi-Lorentzian fits,
however, provide the most uniform description as well as the
possibility to compare the noise components with those found in
neutron star binaries (see Sect.~\ref{sec:disc}).

In almost all of our PSDs, four broad Lorentzians are apparent
(Figs.~\ref{fig:fitex} and~\ref{fig:preflare}) which we call $L_1$,
$L_2$, $L_3$, and $L_4$. The fourth broad component, $L_4$, is usually
present above 32\,Hz and has also been noted by \citet{revnivtsev:00a}
and \citet{nowak:00a}. This component is also seen in other black hole
or neutron star sources \citep{vanstraaten:01a,belloni:02a}.  In the
P30157 data set only its low frequency part can be seen.  However,
with the knowledge from the P40099 and P50110 data sets, we
constrained its shape, especially its approximate width, and included
this component in the P30157 models. We will discuss the issue of
possible systematic errors resulting from this approach later.

We generally achieved a good fit using four broad Lorentzian
components, $L_1$--$L_4$, with typical peak frequencies of
$\sim$0.2\,Hz, $\sim$2\,Hz, $\sim$6\,Hz, and $\sim$40\,Hz.  In our
first modeling attempts we left the $Q$-factors of the Lorentzians as
free parameters. They were found to vary considerably from observation
to observation: $Q_1$, $Q_2$, and $Q_3$ ranged from $\sim$0.1--2.8,
$Q_4$ was even less constrained.  However, later experiments showed
that fixing each $Q$-value to a representative value for each broad
component does not significantly alter the fit results.  This can be
expected, since the broad components overlap over large frequency
intervals (Figs.~\ref{fig:fitex}b and~\ref{fig:preflare}b). We
therefore set $Q_1=0.25$, $Q_2=0.3$, $Q_3=0.3$, and $Q_4=0.5$ for all
fits.  This resulted in a good description of most PSDs.  There are
only 22~cases out of the total 130 PSDs where the fit significantly
improved when one or more of the $Q$-factors were treated as a free
parameter.  In these cases, which mainly coincided with the ``failed
state transitions'' further discussed in Sect.~\ref{sec:failed} below,
we accepted the model with a variable $Q$ as the best fit. For the low
signal to noise \emph{energy resolved} PSDs of the P30157 data we
fixed $Q_4$ and $f_4$ to their best-fit value of the non energy
resolved PSD.

In addition to $L_1$ through $L_4$, a power law of the form
\begin{equation}
\text{PL}(f)=R_{\text{pl}} f^{-\alpha_{\text{pl}}}
\end{equation}
is required in $\sim$40\% of the observations.  The majority of these
power laws are required to reduce the residuals below $\sim$0.01\,Hz.
They have a typical slope of $\sim$1.3 and a normalization of
$\sim$$10^{-4}$.  In some instances the power law is the main
contributor to the total rms amplitude. As we show in
Sect.~\ref{sec:failed}, these observations mainly coincide with
``failed state transitions''.  Generally, these observations are the
same where we also had to leave the $Q$-values of the Lorentzians a
free parameter. This can be expected as the contribution of the power
law at higher frequencies slightly contaminates the other fit
parameters. In other observations -- normally those close to the
``failed state transitions'' -- a clear deviation from the fit with
broad Lorentzians is also present at low frequencies. Due to the
logarithmic binning of our data, in these cases the deviations were
quite often only seen in the lowest few frequency bins and thus there
was not sufficient information to reliably constrain the shape of the
deviation. For consistency with the fits performed for the ``failed
state transitions'', we chose to describe these observation with the
same PSD model that we also used for the intermediate or soft state,
thus implying that the reason for the low-frequency excess in these
observations is the same as that for the excess during the ``failed
state transitions''. Since the uncertainty of the power law slope is
rather large or undeterminable due to numerical reasons for these
latter observations, we do not give specific uncertainties. We also
caution that the difficulty in constraining the power law also results
in an anti-correlation between $\alpha_{\text{pl}}$ and the power law
normalization $R_{\text{pl}}$, which we view as being largely
systematic in nature.

In agreement with \citet{nowak:00a}, \citet{nowak:01b}, and
\citet{belloni:02a}, we often see additional subtle but statistically
significant substructures.  Sometimes these structures are sharp
(Figs.~\ref{fig:fitex} and~\ref{fig:preflare}) or weak and broad. In
this case, further Lorentzian components ($L_{\text{add},i}$) were
added to the basic PSD continuum.  We generally retained these
components if their addition resulted in a significant improvement of
the best fit $\chi^2$, or if they were very clearly present in the
PSDs\footnote{As mentioned in Sect.~\ref{subsec:calc}, the PSDs are
  logarithmically rebinned to give a good description of the continuum
  power. This rebinning often had the result that the narrow features
  were confined to only one or two frequency bins. As a result, their
  modeling with a Lorentzian could not improve the $\chi^2$, although
  the feature is clearly present in the unbinned power spectrum. To be
  at least able to note the presence of such a feature, we
  nevertheless included them in the fit; however, only their frequency
  and power can be constrained, while their width, as represented by
  the $Q$-value, can only be described as ``narrow''.  These features
  will generally have $Q\gtrsim 50$ in Table~\ref{tab:fitresults}.
  Because of these complications, we also do not give any formal
  uncertainty for the narrow Lorentzians in
  Table~\ref{tab:fitresults}.\label{psdfoot}}. Residuals at
frequencies slightly higher than the peak frequency of $L_1$ tended to
be narrow. We modeled these features with one or two narrow
Lorentzians, $L_{\text{add},1}$ and $L_{\text{add},2}$ (see
Fig.~\ref{fig:fitex} and Fig.~\ref{fig:preflare}).  Typical peak
frequencies for these QPOs range from 0.1 to 0.6\,Hz, and one might
speculate about their relationship to the horizontal branch
oscillation peaks in neutron star X-ray binaries
\citep[][Table~4]{wijnands:99a}.  In addition, below $L_1$ a broad,
weak ($\lesssim2\sigma$) residual at $\sim$0.03\,Hz was often seen,
which we designate $L_{\text{add},3}$. The total contribution of
$L_{\text{add},1}$, $L_{\text{add},2}$, and $L_{\text{add},3}$ to the
rms is on the order of a few percent at most. See
Sect.~\ref{sec:thin} for a discussion of the behavior of these
additional components.

Using this modeling approach, we obtained values for the reduced
$\chi^2$ in the range of 1--2. The best-fit parameters of all 130
observations are given in electronic form in
Table~\ref{tab:fitresults}.  Most of the remaining systematic residuals
in these fits are probably due to the fact that the broad noise 
components have a more complex shape than a simple Lorentzian profile
and/or that sub-harmonics are present. This is especially true for
$L_2$, which is often associated with $\lesssim$$2\sigma$ residuals in
the $\sim1$--$2$\,Hz range \citep[see also][]{nowak:00a}. The broad
low-frequency residual $L_{\text{add},3}$ and other residuals at
frequencies below $\nu_1$ might also be due to this effect. A further
source of weak systematic residuals might be our choice of fixed
$Q$-factors.

After obtaining the final best fit we computed the total rms
variability amplitude relative to the mean count rate by summing the
contributions of all power spectral fit components. The best fit
Lorentzian profiles were integrated from zero to infinity
(Eq.~\eqref{eq:rms}), while the power law contribution was obtained by
integrating the power law over the frequency range from $10^{-3}$\,Hz
to 200\,Hz.  Because of the different time resolution of the extracted
lightcurves this approach is more systematic than the direct
measurement of the rms variability from the X-ray data.

\subsection{Time Lags, Coherence, and Spectral Modeling}
Although the emphasis of this paper is on the evolution of the short
term timing properties of Cyg~X-1 in terms of the PSD, it is obviously
important to also characterize the source in terms of other
quantities.

For black hole candidates, the X-ray lightcurves in two energy bands
are similar to each other. However, the X-ray lightcurve in a higher
energy band generally lags that measured in a lower energy band. This
X-ray time lag depends on the Fourier frequency. A measure for the
similarity of the X-ray lightcurves in these two bands is given by
their coherence, which is again Fourier frequency dependent. We
compute both quantities using the formulae given by \citet{nowak:98b}.
As we have shown previously, an especially interesting quantity is the
average X-ray time lag in the 3.2--10\,Hz band
\citep{pottschmidt:00a}, which we will use to quantify the X-ray lag.

To obtain a description of the X-ray photon spectrum, we describe each
source spectrum as the sum of a power law spectrum with photon index
$\Gamma$ and a multi-temperature disk-black body after
\citet{makishima:86a}, characterized by the temperature at the inner
edge of the accretion disk, $kT_{\text{in}}$. To this continuum, a
reflection spectrum after \citet{magdziarz:95a} is added.  For this
paper we concentrate on the X-ray spectrum below $\sim$20\,keV by
ignoring the HEXTE data.  Such an empirical model is sufficient to
roughly describe the most important spectral parameters.  A full study
of the evolution of the broad-band X-ray spectrum that includes the
HEXTE data and models the source spectrum in terms of the
Comptonization models of \citet{dove:97b}, \citet{coppi:99a}, and
\citet{poutanen:96b} will be presented in a forthcoming paper
\citep{gleiss:02a}.

\section{Evolution of the Power Spectrum}\label{sec:evolution}

\begin{figure*}
\centering
\includegraphics[width=17cm]{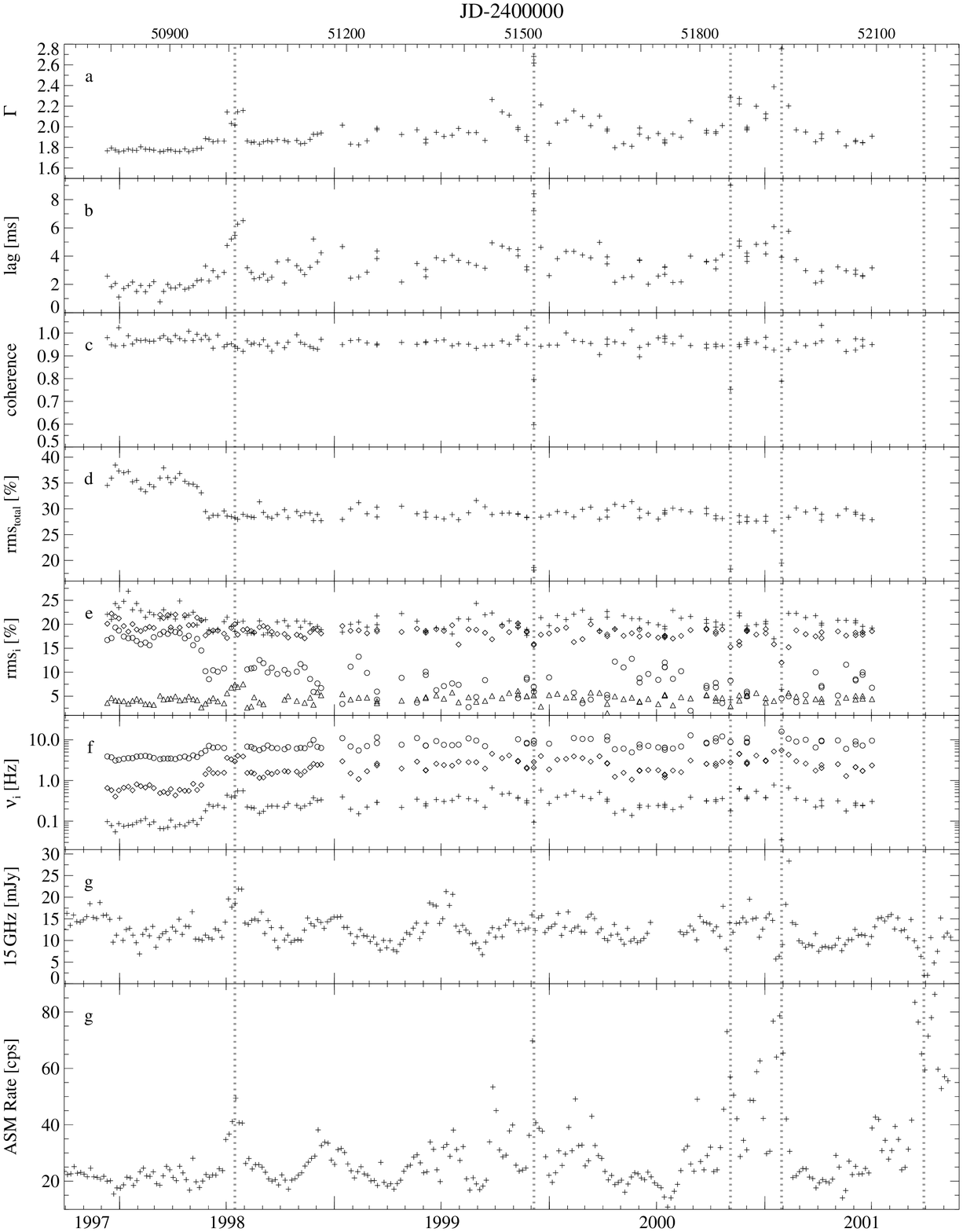}
\caption{Spectral and timing parameters of Cyg~X-1 between 1997 and 2001.
\textbf{a)} Photon index.
\textbf{b)} X-ray time lag between 2--4\,keV and 8--13\,keV,
  averaged over the 3.2--10\,Hz frequency band.
\textbf{c)} X-ray coherence between 2--4\,keV and 8--13\,keV,
  averaged over the 3.2--10\,Hz frequency band.
\textbf{d)} Relative rms amplitude for
  the total 2--13.1\,keV power spectrum.
\textbf{e)} Relative rms amplitude and \textbf{f)} frequency 
  of each of the four broad noise components. Crosses:
  $\text{rms}_1$, diamonds: $\text{rms}_2$, circles: $\text{rms}_3$,
  triangles: $\text{rms}_4$.
\textbf{g)} Daily averaged 15\,GHz radio flux, determined with the Ryle
   telescope.   
\textbf{h)}:  RXTE ASM 2--12\,keV count rate.
The dotted vertical lines mark special observations discussed in
Sect.~\ref{sec:failed} and Sect.~\ref{sec:02soft}. The radio flux and
the ASM rate have been rebinned to a resolution of $\sim$5.6\,d, the
orbital period of Cyg~X-1, for clarity. }
\label{fig:rms}
\end{figure*}

\begin{figure}
\resizebox{\hsize}{!}{\includegraphics{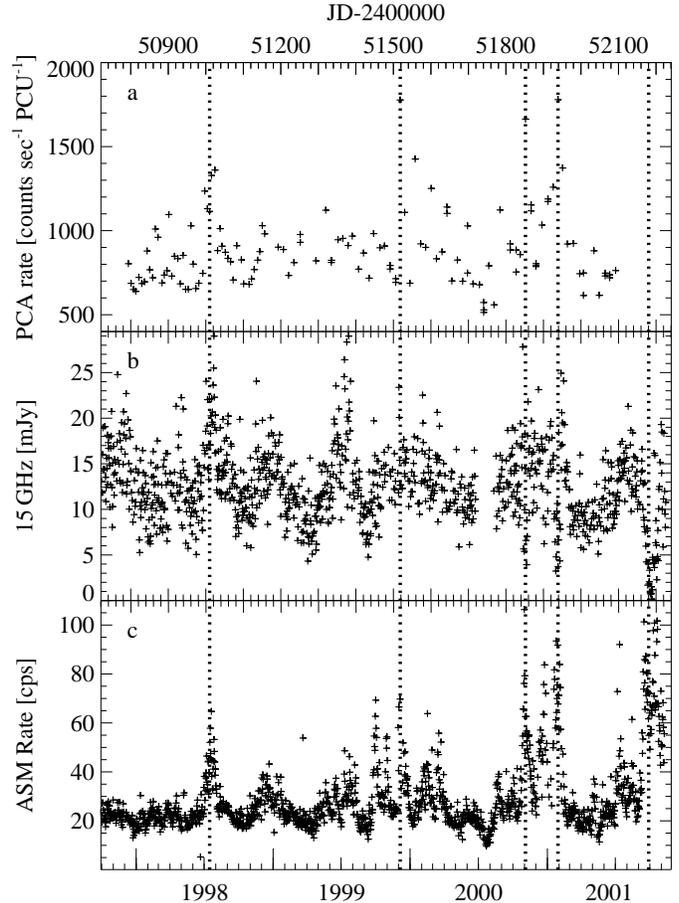}}
\caption{\textbf{a)} RXTE-PCA count rate (2.5--20\,keV) for each
  pointing, as well as \textbf{b)} RXTE-ASM count rate and \textbf{c)}
  15\,GHz radio flux, both binned to a resolution of 1\,d to show the
  variability on timescales that are smaller than the orbital
  timescale.}
\label{fig:rates}
\end{figure}

\begin{table*}
\caption{Average values and standard deviations of the rms amplitudes,
  $\text{rms}_i$, the peak frequencies, $\nu_i$, and the frequency
  ratios, $\nu_i/\nu_j$, for the four broad noise components
  ($i,j=1, \ldots, 4$). The values are derived for different time
  episodes surrounding the 1998~May change and the 1998~July failed
  state transition. For the P30157 data the highest peak frequency, $\nu_4$,
  is biased (see text) and is therefore not quoted in this table.}   
\label{tab:averages} 
\noindent\begin{tabular}{cccccc} 
    & P30157         & P30157    & P30157   & P30157   & P40099/P50110$^b$\\ 
    & (1--18)$^a$ & (19--28)     & (29--32) & (33--52)       & \\ 
    & before 1998~April & 1998~May/June  & 1998~July flare &  post 1998~July & 1999--2001\\  
\hline
$\text{rms}_{\text{tot}}$ [\%] &
                36$\pm$1 & 32$\pm$3 & 28.2$\pm$0.2 & 29$\pm$1 & 29$\pm$2\\ 
$\text{rms}_1$ [\%] &
                23$\pm$2 & 20$\pm$1 & 20$\pm$1 & 19$\pm$4 & 20$\pm$4\\ 
$\text{rms}_2$ [\%] &
                20$\pm$1 & 20$\pm$1 & 19$\pm$1 & 18$\pm$1 & 18$\pm$1\\ 
$\text{rms}_3$ [\%] &
                17$\pm$1 & 13$\pm$3 & -- & 10$\pm$2 &  6$\pm$4\\ 
$\text{rms}_4$ [\%] &
                 4$\pm$1 &  4$\pm$3 &  7$\pm$1 &  3$\pm$2  &  5$\pm$1\\ 
$\nu_1$ [Hz] & 
              0.08$\pm$0.02 & 0.16$\pm$0.07 & 0.45$\pm$0.07 & 0.25$\pm$0.06 & 0.33$\pm$0.12\\ 
$\nu_2$ [Hz] & 
              0.6$\pm$0.1 & 1.1$\pm$0.5 & 3.5$\pm$0.5 & 1.7$\pm$0.4 & 2.6$\pm$0.9\\
$\nu_3$ [Hz] & 
              3.6$\pm$0.3 & 5.2$\pm$1.3 & -- & 6.4$\pm$1.0 & 8.4$\pm$2.0\\
$\nu_4$ [Hz] & 
                  &     &     &     & 41$\pm$8\\
$\nu_2/\nu_1$ & 
              7.3$\pm$1.1 & 7.0$\pm$0.7 & 7.7$\pm$0.7 & 7.0$\pm$0.5 & 7.8$\pm$1.3\\ 
$\nu_3/\nu_1$ & 
              44.0$\pm$6.4 & 35.6$\pm$8.4 & --  & 27.7$\pm$4.5 & 30.0$\pm$5.3\\
$\nu_4/\nu_1$ & 
                  &     &     &     & 141$\pm$65\\
$\nu_3/\nu_2$ & 
              6.0$\pm$0.9 & 5.1$\pm$1.2 &    & 4.0$\pm$0.7 & 3.9$\pm$0.6\\
$\nu_4/\nu_2$ & 
                  &     &     &     & 18.2$\pm$7.2\\
$\nu_4/\nu_3$ & 
                  &     &     &     & 5.1$\pm$1.4\\
\hline
\end{tabular}

\footnotesize

${}^{a}$ observation number within the P30157 data set\\  
${}^{b}$ the averages for P40099 and P50110 include several failed state
transitions.
\end{table*}

\begin{figure}
\resizebox{\hsize}{!}{\includegraphics{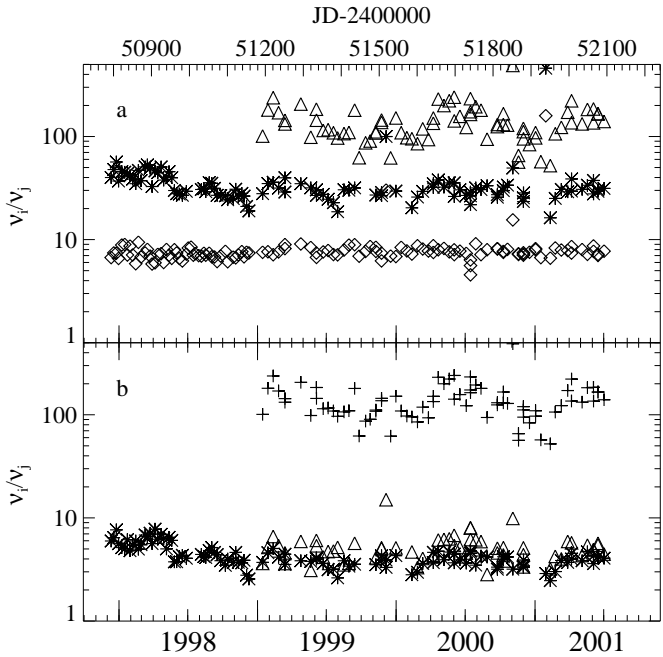}}
\caption{Constancy of the peak frequency ratios over the monitoring
  campaign. \textbf{a)} Diamonds: $\nu_2/\nu_1$, stars: $\nu_3/\nu_1$,
  triangles: $\nu_4/\nu_1$. \textbf{b)} Stars: $\nu_3/\nu_2$, triangles:
  $\nu_4/\nu_2$, crosses: $\nu_4/\nu_3$.}
\label{fig:ratio}
\end{figure}

In this section we will discuss the overall behavior of the PSD of
Cyg~X-1 since 1998. The evolution of the best fit PSD parameters, the
X-ray time lag, and the X-ray coherence function, as
well as the photon index, $\Gamma$, from the spectral fits, the
15\,GHz radio flux, and the RXTE ASM count rate are shown in
Fig.~\ref{fig:rms}. Both the radio flux and the ASM count rate have
been rebinned to a resolution corresponding to the orbital period of
the HDE~226868/Cyg~X-1 system \cite[$P_{\text{orb}}=5.599829(16)$\,d,
see][]{brocksopp:98a,lasala:98a}. This rebinning removes the effect of 
the well known orbital variation of the X-ray and radio flux
\citep{pooley:98a,brocksopp:99a,wen:01a} such that long term changes
become more apparent. Due to the rebinning, however, many short flares
that are not due to the orbital modulation are smoothed out. We
therefore present the radio and ASM data again in Fig.~\ref{fig:rates}
with a higher time resolution, together with the RXTE PCA count rate
determined from the monitoring observations. Not unexpectedly, the ASM
and PCA rates are well correlated, although their correlation is not
perfect due to their different energy bands and the X-ray spectral
variability.

\subsection{The Two Hard States}\label{sec:twohard}
The most striking result of our fits is the bimodal behavior of the
total rms amplitude (Fig.~\ref{fig:rms}d): Starting in 1998 mid-April,
the overall variability of Cyg~X-1 on short timescales showed a
decrease of $\sim$7\%, from an average rms amplitude of 36$\pm$1\% to
29$\pm$1\%.  At the same time, the power law index of the X-ray
spectrum slightly softened from $\sim 1.75$ to $\sim 1.85$, the
characteristic frequencies of $L_1$ through $L_3$ shifted to higher
values, and the X-ray time lag increased. Furthermore, a larger
fraction of observations requires the power law component below
$\sim$0.01\,Hz.  Apart from further ``X-ray flares'', which we will
describe in Sect.~\ref{sec:failed}, Cyg~X-1 has remained in this
softer and less variable (on short time-scales) X-ray state.
Table~\ref{tab:averages} summarizes the main properties of the PSD
before, during, and after the 1998 changes.

The parameters for the X-ray spectrum and timing behavior measured
before 1998~May are very similar to the canonical hard state values
for this source. Our campaign shows that these values have not been
reached since then, i.e., \emph{since 1998~May, Cyg~X-1 is exhibiting
  a behavior that is atypical compared to previously observed hard
  states of this source.} In the following we describe these changes
in further detail.

The change of the total rms can be mainly attributed to the behavior
of $L_3$ (Fig.~\ref{fig:rms}e): while the rms values of $L_1$ and
$L_2$ stay roughly constant, $L_3$ declines from its pre-May strength
of $\sim 17$\% to 10$\pm$3\%, i.e., almost half of its original
strength\footnote{Due to the high frequency cutoff of the P30157 
  data, the parameters of $L_4$ are difficult to constrain during
  1998. We found no indication for a bimodal behavior of this
  component, however, the large uncertainty of the fit parameters does
  not allow us to make any definitive statements.}.  Especially during
the X-ray flare in 1998~July, $L_3$ cannot be detected at all.  Since
these events, $L_3$ has mainly been present at a level between
$\sim$5\% and $\sim$10\% rms, and was sometimes not detected at all.
Typical examples for pre- and post-1998~May power spectra are shown in
Figs.~\ref{fig:preflare} and~\ref{fig:fitex}, respectively.  These
figures clearly show the decrease in strength of $L_3$, especially in
the panels where the PSD has been multiplied with frequency.

In addition to this large change in strength, the whole PSD shifts to
higher frequencies (Table~\ref{tab:averages}). The two lowest peak
frequencies, $\nu_1$ and $\nu_2$ increase by a factor of $\sim$3
during the change, while $\nu_3$ increases by a factor of $\sim$2.
This behavior is not mirrored by the peak frequency of $L_4$. The
sharp drop from about 60\,Hz to 40\,Hz at the end of P30157 in $\nu_4$ is
due to the biasing of its value caused by the low frequency cutoff of
our data. We note that although the peak frequencies change, their
ratios are remarkably constant over the whole campaign -- in fact, the
1998~May event is only barely discernible in the evolution of the
frequency ratios (Fig.~\ref{fig:ratio}).

\begin{figure}
\resizebox{\hsize}{!}{\includegraphics{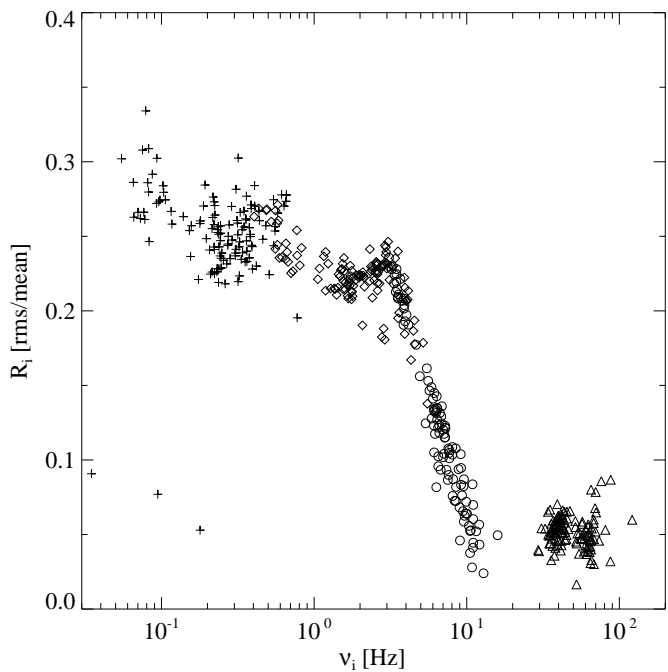}}
\caption{Normalization  of the four broad noise components, $R_i$,
  versus their peak frequencies, $\nu_i$. The different plot symbols
  are attributed to the same components as in Fig.~\ref{fig:rms}.}
\label{fig:norm}
\end{figure}

The nature of the change in the hard state timing behavior can be
clearly seen when the variations in the variability amplitudes and
peak frequencies are considered together.  Fig.~\ref{fig:norm} shows
the dependence of the normalization constant, $R_i$, of each broad
component on the associated peak frequencies, $\nu_i$, for our data.
Regions in Fig.~\ref{fig:norm} where the data point density is higher
are due to the hard state behavior before and after 1998~April.

Several interesting features are apparent in Fig.~\ref{fig:norm}.  The
three lowest frequencies, $\nu_1$ through $\nu_3$, form a continuum
from about $5\,10^{-2}$\,Hz to 20\,Hz. For all three Lorentzians, the
normalization of the Lorentzian correlates with its peak frequency --
higher $\nu_i$ also imply a lower normalization. Furthermore, in those
frequency regions, where several Lorentzians can be found, e.g.,
around 0.5\,Hz for $\nu_1$ and $\nu_2$, and around 5\,Hz for $\nu_2$
and $\nu_3$, their normalizations seem to be the same. We stress that
this is not due to a misidentification of the Lorentzians as these are
always clearly distinguishable in the observations presented here.
There are two distinct regions of the $R_i$-$\nu_i$ correlation: below
$\sim 3$\,Hz, where the figure is dominated by $L_1$ and $L_2$, and
above 3\,Hz, where $L_3$ dominates. Clearly, $\nu_3$ and $R_3$ are
much more strongly correlated than the parameters for the low
frequency Lorentzians.  We note that the sharpness of this correlation
might in principle have been influenced by the 32\,Hz maximum
frequency of our 1998 data set.  However, the strong $R_3$-$\nu_3$
correlation is also present when only considering the P40099 and
P50110 data.  We speculate that these correlations are similar to the
effect first seen by \citet{belloni:90a} in terms of their broken
power law analysis of the PSDs of Cyg~X-1, where the break frequency
of the PSD is correlated with the total source rms variability. In our
Lorentzian decomposition, $\nu_1$ is roughly equivalent to the break
frequency and shows a similar behavior. What is new, and not yet
understood, however, is the behavior of $L_3$.

Finally, the highest peak frequency component, $L_4$ (triangles), is
clearly distinct from the lower frequency features and shows a
different behavior. Its peak frequency and normalization are
comparatively stable. The clustering of the data at 40 and 60\,Hz
again reflects the fact that $\nu_4$ has been systematically
over-estimated during P30157. During P40099 and later, however, no
systematic changes are seen in $\nu_4$, although our data should have
been sensitive to such changes.

\begin{figure}
\resizebox{\hsize}{!}{\includegraphics{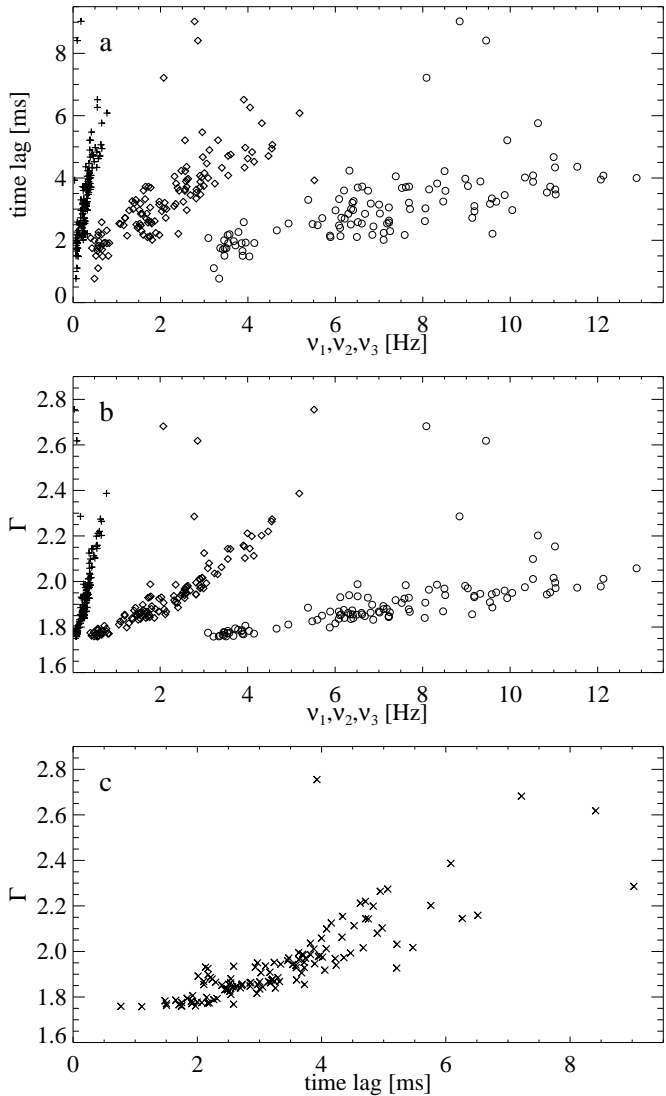}}
\caption{Correlations  \textbf{a)} between the average time lag in the
  band between 2--4\,keV and 8--13\,keV and the peak frequencies
  $\nu_1$, $\nu_2$, and $\nu_3$ of the lower three Lorentzians,
  \textbf{b)} between the photon index, $\Gamma$, of the X-ray spectrum
  and $\nu_1$ through $\nu_3$, and \textbf{c)} between $\Gamma$ and the
  X-ray time lag. The symbols in panels~a and~b correspond to those
  used in Fig.~\ref{fig:rms}.}
\label{fig:multicorr}
\end{figure}

\begin{figure}
\resizebox{\hsize}{!}{\includegraphics{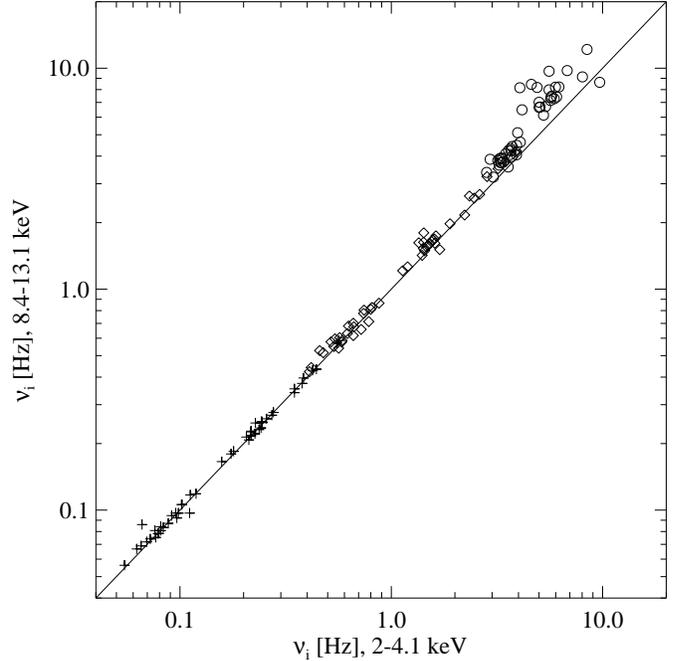}}
\caption{Peak frequencies of $L_1$, $L_2$, and $L_3$ as determined for
  the energy band from $\sim 2$--4.6\,keV versus the peak frequency
  for the energy band from 8.4--13.1\,keV. The symbols correspond to
  those used in Fig.~\ref{fig:rms}. } 
\label{fig:psd_specevol}
\end{figure}

We now discuss the properties of the Lorentzians in terms of other
measured quantities (Fig.~\ref{fig:multicorr}).  The distinct behavior
of the Lorentzians when plotted against the X-ray photon spectral
index, $\Gamma$, and the X-ray time lag gives us further confidence in
the individual identifications of the Lorentzians. Both the X-ray time
lag and the peak frequencies $\nu_1$ through $\nu_3$ are clearly
correlated with $\Gamma$: softer photon spectra imply a shift of the
PSD towards higher frequencies and also higher X-ray time lags
(Fig.~\ref{fig:multicorr}b and~c). There is also a clear correlation
between the average time lag and the Lorentzian peak frequencies
(Fig.~\ref{fig:multicorr}a). Although a correlation between photon
index and frequency has been noted before with more limited data sets
\citep{dimatteo:99a,gilfanov:99a,revnivtsev:00a,nowak:01b}, this is
the first time that a simultaneous correlation with the time lags has
been found for Cyg~X-1. A similar behavior has been seen in a more
limited hard state data set of GX~339$-$4 \citep{nowak:01b}.
Recently, a comparative study of the ``QPO--$\Gamma$'' correlations
has been performed for several BHCs with narrow QPOs by
\citet{vignarca:03a}. While the direct identification of one of our
correlations of Fig.~\ref{fig:multicorr}b with the unified track
displayed in their work is not easily possible (different spectral
models were used to derive $\Gamma$), the parameter space is
comparable. We do not see any turnoff, i.e., no decreasing photon
index at high frequencies which seems to be a common feature in some
of the sources studied by \citet{vignarca:03a} (mainly GRS~1915$+$105,
GRO~J1655$-$40, and XTE~J1550$-564$). This is consistent with the
picture suggested by those authors that the turnoff might be related
to a transition between the hard and the very high state, since the
latter is not observed in Cyg~X-1 (at least not in its canonical form,
note, however, that the hard spectral component is generally present
also in the soft state).

We note that there is no strong energy dependency of the power
spectrum shape during the normal hard state phases. As an example,
Fig.~\ref{fig:psd_specevol} shows the correlation between the peak
frequency of $L_1$, $L_2$, and $L_3$ for the two energy bands for the
1998 data.  To within their error bars, the peak frequencies of $L_1$
and $L_2$ are identical for these bands. The peak frequency of $L_3$
is higher by about 30\% in the hard energy band, a trend that
continues when analyzing lightcurves in even harder energy
bands. This trend is similar to that seen in the Lorentzian
decomposition of the PSD of \object{XTE J1650$-$500}
\citep{kalemci:02a}.

\begin{figure*}
\includegraphics[width=12cm]{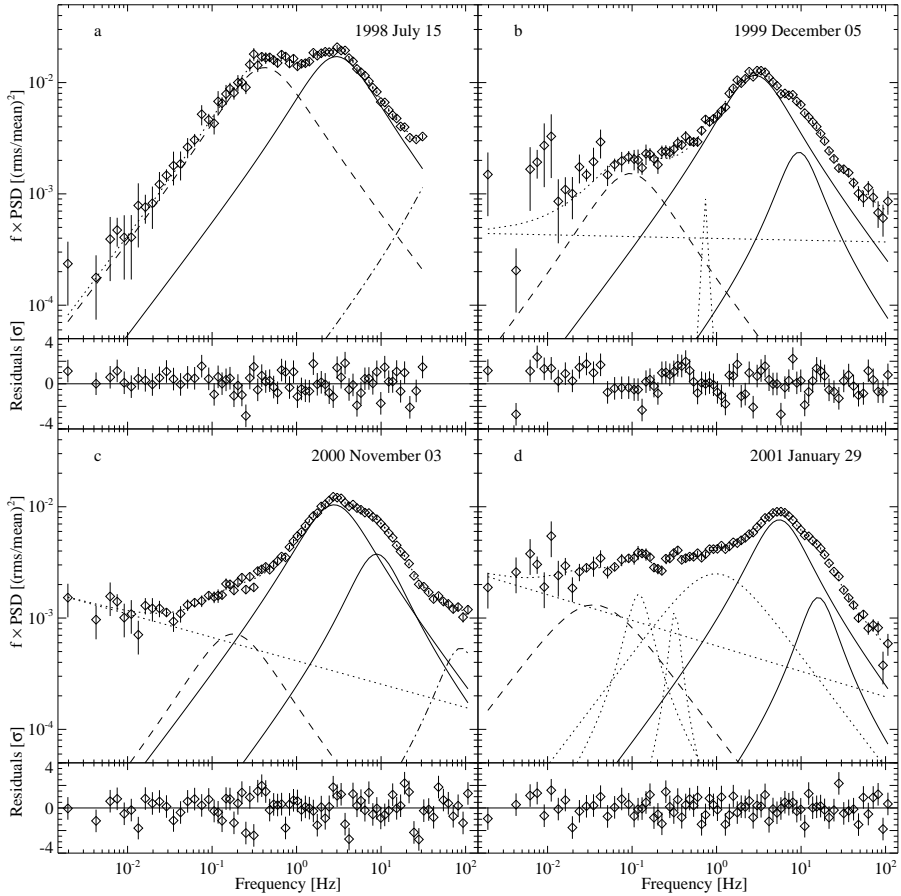}

\caption{\textbf{a--d)} Examples of power spectra that deviate
  significantly from the typical hard state shape as defined by
  Figs.~\ref{fig:fitex} and~\ref{fig:preflare}. The best fit broad
  noise components are distinguished by their line styles: dashed:
  $L_1$, solid: $L_2$ and $L_3$, dash-dotted: $L_4$. The remaining
  components and the total best fit are dotted.  The observation dates
  are given in the plot, they are also marked by dotted lines in
  Figs.~\ref{fig:rms} and~\ref{fig:rates}.}
\label{fig:failed}
\end{figure*}

\subsection{Failed State Transitions}\label{sec:failed} 
In this section we study the behavior of Cyg~X-1 during the ``flares''
apparent in the ASM lightcurve of Fig.~\ref{fig:rms}. During these
flares, the rms amplitude decreases, the strength of the power law
contribution relative to the Lorentzians increases, and the X-ray
spectrum softens.  Generally, the X-ray time lag also shows values
that are much higher than those seen before and after the flare, and
the coherence function drops. PSDs measured during several of these
events are shown in Fig.~\ref{fig:failed}.  These examples are
significantly different from the ``standard PSD shape'' for the hard
state as defined in Figs.~\ref{fig:fitex} and~\ref{fig:preflare}.

Crucial to the interpretation of these flares is the evolution of the
PSD and the other timing quantities over the flares. We will
concentrate on the flares best sampled by our monitoring observations.
These flares, which were observed in 1998~July, 1999~December, and
2000~November, are identified by dashed lines in Fig.~\ref{fig:rms}.
Additional smaller events were also seen, e.g., in 1999~September, in
2000~March, and in 2000~December.  For these latter events, however,
our observations did not sample the change of the PSD in sufficient
detail.

\begin{figure*}[t]

\includegraphics[width=12cm]{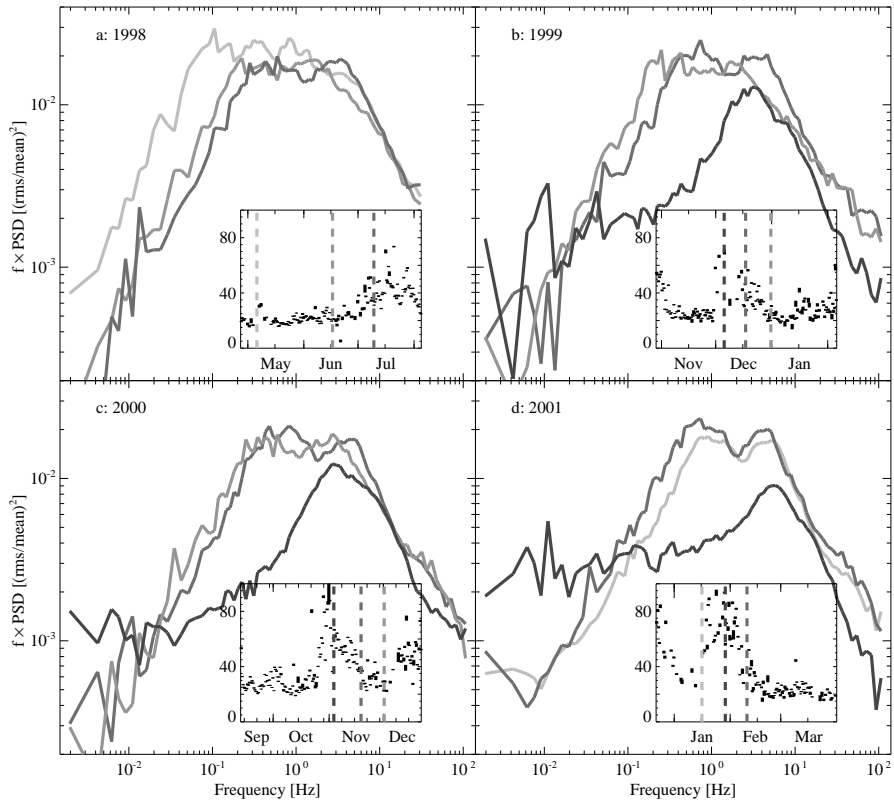}
\caption{Development of the power spectra near the flaring events seen
  in the ASM lightcurve. Their location in the ASM lightcurve is
  indicated in the insets. The panels show the data from \textbf{a)}
  1998 July, \textbf{b)} 1999 December, \textbf{c)} 2000
  October/November, and \textbf{d)} 2001 January/February.}
\label{fig:seq}
\end{figure*}

A typical example for the evolution during the first stage of these
flares was seen in 1998~July (Figs.~\ref{fig:failed}
and~\ref{fig:seq}): Here, the characteristic frequencies of the
Lorentzians shift to higher frequencies and their relative strength
changes.  In contrast to the standard PSD, the $L_3$ component is weak
or missing and the peak frequencies of $L_1$ and $L_2$ are
significantly enhanced.  Furthermore, the X-ray time lag in the
3.2--10\,Hz band increases (Fig.~\ref{fig:rms}b, see also
\citealt{pottschmidt:00a}).  On the resolution of our monitoring, the
transition back into the hard state mirrors that of the transition
into the flare.

\begin{figure}
  \resizebox{\hsize}{!}{\includegraphics{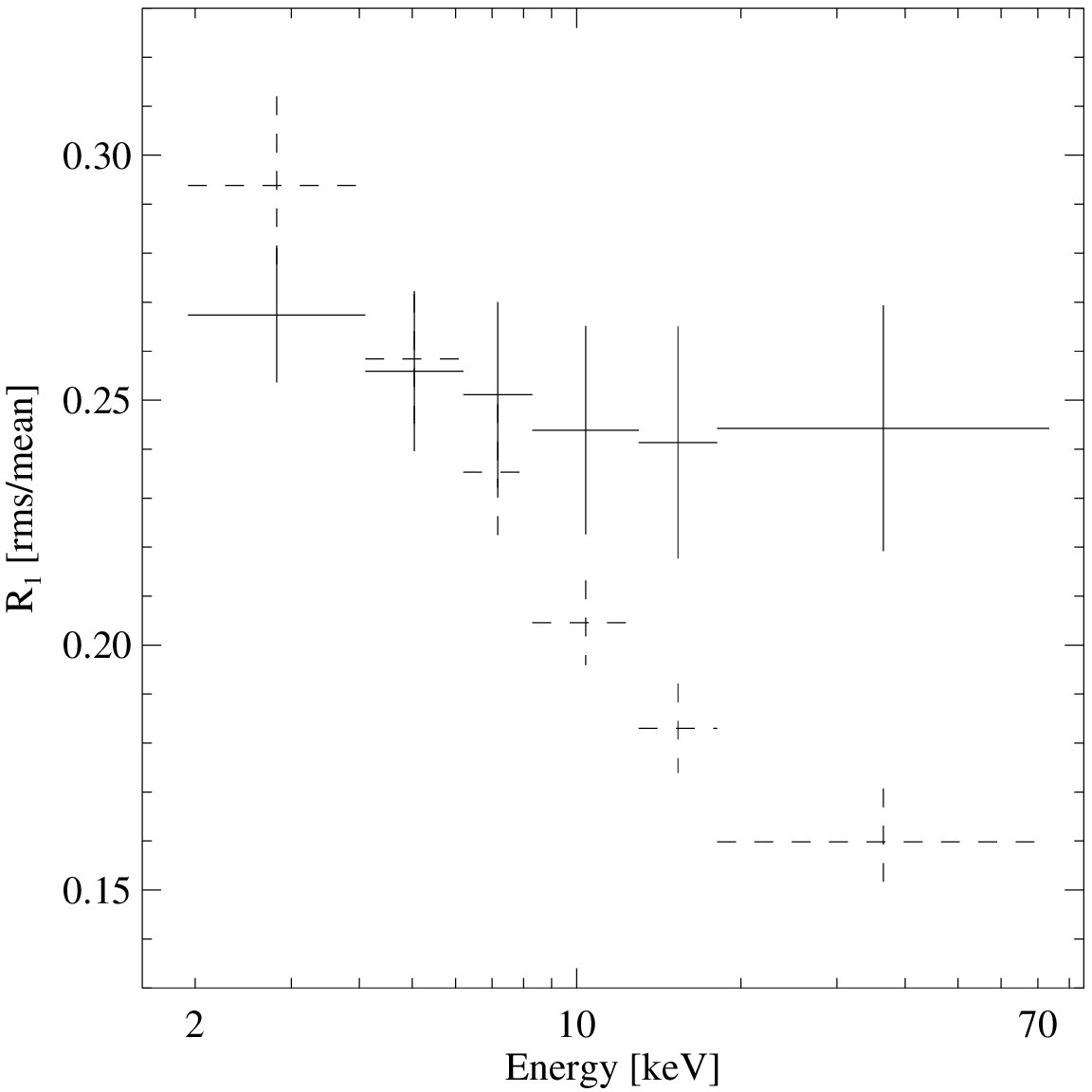}}
\caption{Change of normalization of $L_1$ as a function of energy
  during the first half of 1998. Solid lines: Average of the normalization
  of $L_1$, $R_1$, for observations P30157/19 through 28, \emph{before} the
  1998~July failed state transition. Dashed lines: Average of $R_1$ for
  observations P30157/29 through 33, \emph{during} the 1998~July
  event.}\label{fig:psd_normcorr} 
\end{figure}

We briefly note that during the 1998~July event there is a change in
the energy dependency of the rms variability amplitude
(Fig.~\ref{fig:psd_normcorr}).  Before the flare, the rms variability
amplitude of $L_3$ increased with photon energy, that is the PSD was
flatter at higher energy bands as is typical for the hard state
\citep{nowak:98b}.  During the flare, when the spectrum was soft, the
rms variability amplitude decreased with energy.  This is best
illustrated in the behavior of $L_1$, where the total power in this
component before the 1998~July event has a rather small energy
dependence (Fig.~\ref{fig:psd_normcorr}, solid lines), while it is
strongly energy dependent during the flare
(Fig.~\ref{fig:psd_normcorr}, dashed lines).

\begin{figure}
  \resizebox{\hsize}{!}{\includegraphics{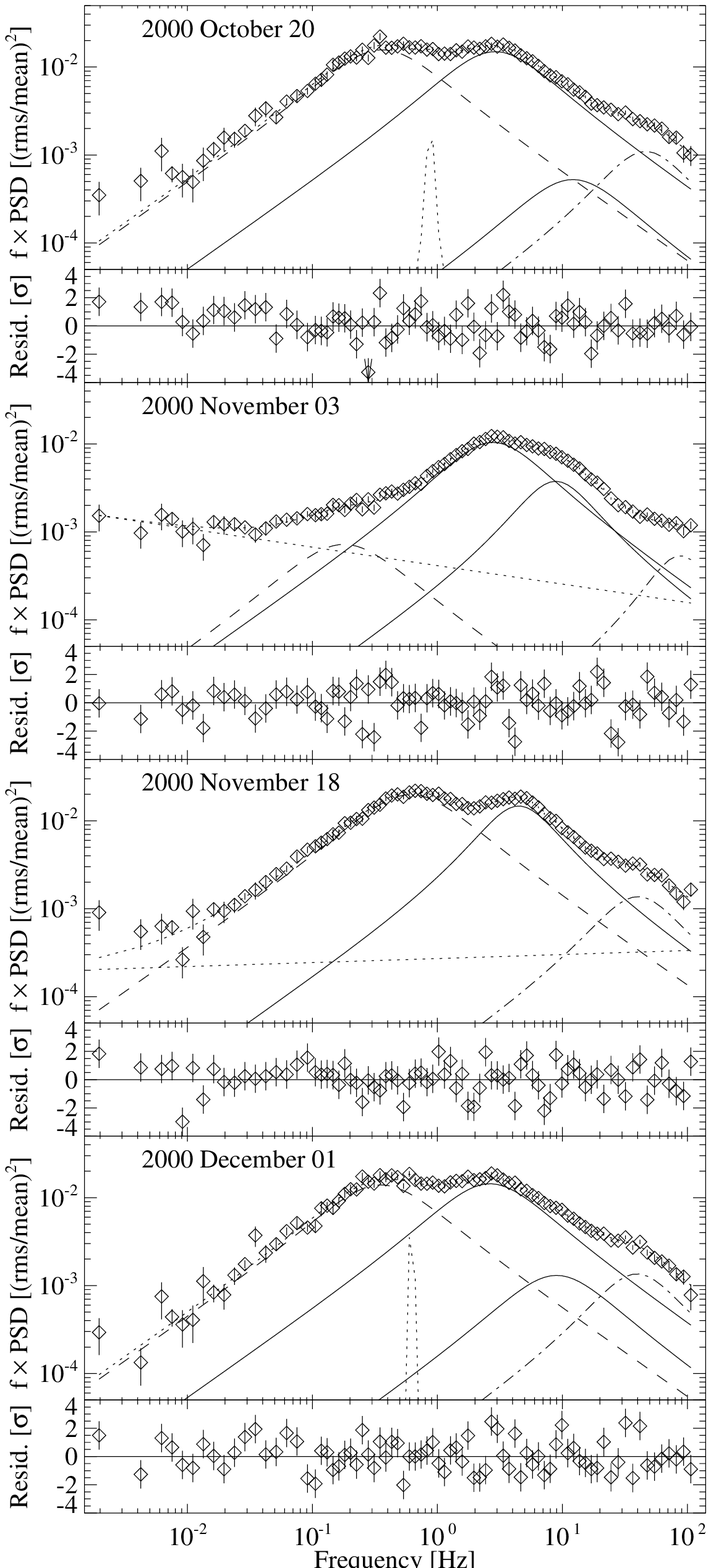}}
\caption{Evolution of the PSD during the monitoring observations of
  2000~November. The line style of the best fit components is the same
  as that in Fig.~\ref{fig:failed}.}\label{fig:transseq}
\end{figure}

During some of the flares the PSD evolution starts similar to
1998~July but then continues until a strong low-freqency power law
dominates the observed rms, instead of the Lorentzians.  Typical
examples for such a behavior are observations made on 1999 December~05
and on 2000 November~03, at the peak of their respective outbursts
(Figs.~\ref{fig:seq}b and~c). Of the Lorentzians, $L_2$ is strong and
$L_3$ is clearly present, albeit at the weak levels typical of PSDs
observed after 1998~May.  On the other hand, $L_1$ is barely
identifiable.  Both observations also have quite large X-ray lags
(e.g., $\sim$8\,ms for 2000 November) and decreased coherence (e.g.,
0.6 on 1999 December~05). The neighboring observations, however,
display the characteristic double-peaked PSD similar to 1998~July,
coherence close to unity, and lower lags of $\sim$3\,ms. The evolution
of the Lorentzians through the 2000 November flare is shown in
Fig.~\ref{fig:transseq}. Both, the PSD shape as well as the X-ray
spectral parameters seen here are very similar to earlier observations
of Cyg~X-1 reported by \citet{belloni:96}. Comparing the timing
properties and spectrum of \object{GS~1124$-$683} \citep[Nova Muscae
1991;][]{ebisawa:94,miyamoto:94} and GX~339$-$4 \citep{mendez:97} to
those of Cyg~X-1, \citet{belloni:96} found these to be very similar to
the ``intermediate state'' seen in these X-ray transients. In
transients, the ``intermediate state'' is typically seen during
transitions between the soft and the hard state, which fits the
picture drawn above from our RXTE monitoring.

Based on our observations of increased X-ray time lags during the
\emph{transitional phases} between the 1996 hard and soft state and
during the ASM flares, we have previously called the ASM flares
``failed state transitions'' \citep{pottschmidt:00a}. As outlined
above, the behavior of the other timing quantities apart from the
X-ray time lags now justifies our use of this term -- during the
``failed state transitions'', the source changes its behavior compared
to the normal hard state, by sometimes even entering the intermediate
state, but a full soft state is generally not reached, at least not on
the sampling timescale of our RXTE monitoring. 

The availability of the different timing quantities for our monitoring
observations also allows us to connect the increased X-ray time lag
with the observed changes in the power spectrum and the coherence
function.  As we mentioned above, the X-ray time lag varies strongest
in the 3.2--10\,Hz band \citep{pottschmidt:00a}.  The main
contributions to the total rms variability in this frequency range are
due to $L_2$ and, to a lesser extent, due to $L_3$. These two
components are also the only broad noise components that are clearly
present in the intermediate states of 1999 December~05 and 2000
November~03 power spectra (Fig.~\ref{fig:failed}b and~d), as well as
during the transition to the soft state in 1996 (Fig.~2 of
\citealt{pottschmidt:00b}; see also \citealt{belloni:96}).  Since the
time lag behavior does not seem to reflect the transient behavior of
the $L_3$ component, and since the frequency range showing the
enhanced lags is more consistent with that of $L_2$, we tentatively
identify the enhanced time lag during failed state transitions as
being mainly due to $L_2$. Since the soft state itself shows neither
these broad noise components nor enhanced time lags, our
identification provides a self-consistent picture.

While most of the ASM flares are indeed ``failed state transitions'',
there is one flare during the time period covered here, in
2001~January, where the state transition did not fail, but where a
soft state was reached for a brief time.  The evolution of the PSD
shown in Fig.~\ref{fig:seq}d displays again the vanishing of $L_1$ and
$L_4$, with an ``intermediate state'' PSD observed on 2001 January~29.
From the point of view of our monitoring campaign, this behavior would
have led us to classify this flare as yet another failed state
transition. In addition to our RXTE monitoring, however, further
pointed RXTE observations were performed in early 2001, that had a
much better sampling than our campaign. The analysis of these data by
\citet{cui:02a} shows PSDs that are similar to those seen here, with
the exception of the data taken on 2001 January~28. Both the timing
and spectral behavior are very similar to the 1996 soft state, leading
\citet{cui:02a} to claim a possible short soft state of Cyg~X-1. 

We note that in addition to the X-ray properties, the radio flux was
also peculiar in 2001~January. During the end of 2001~January, the
radio flux first dropped to about 30\% of its typical value of
$\sim$15\,mJy, before a radio flare was seen (Figs.~\ref{fig:rms}g
and~\ref{fig:rates}). In contrast to 1996, however, Cyg~X-1 was always
detected in the radio, indicating that the system had not settled into
the soft state as it did in 1996. In fact, our monitoring observation
of 2001 January~29 shows again the transitional PSD. As in
2001~November (Sect.~\ref{sec:02soft}), therefore, the switch from the
soft state (if it can truly be labeled a soft state) back into the
intermediate state occurred within one day.

\subsection{The 2002 Soft State}\label{sec:02soft}
Our RXTE observations performed since 2001~September give further
credence to the identification of the presence of an ``intermediate
state'' during the transitions from the hard to the soft state. This
time, from 2001~September until 2002~October, was the first time since
1996 that Cyg~X-1 exhibited again a full soft state behavior in the
X-rays \emph{and} in the radio emission. A detailed description of
this soft state will be the subject of a later publication in this
series \citep{wilms:03a}, here we will only briefly summarize the
observations of the transition into the soft state during the end of
2001 as far as they are connected with the interpretation of the
flares with the ``intermediate state'' (Sect.~\ref{sec:failed}).

\begin{figure}
\resizebox{\hsize}{!}{\includegraphics{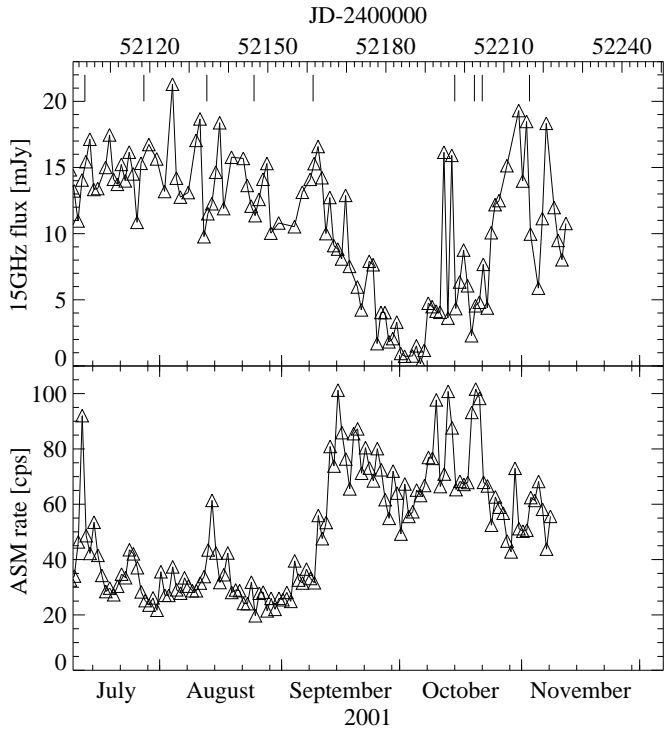}}
\caption{RXTE ASM count rate and 15\,GHz flux of Cyg~X-1 during the
  fall of 2001. The dashes denote the time of those pointed RXTE
  observations that were part of our monitoring
  campaign.}\label{fig:octlc}
\end{figure}

\begin{figure}
  \resizebox{\hsize}{!}{\includegraphics{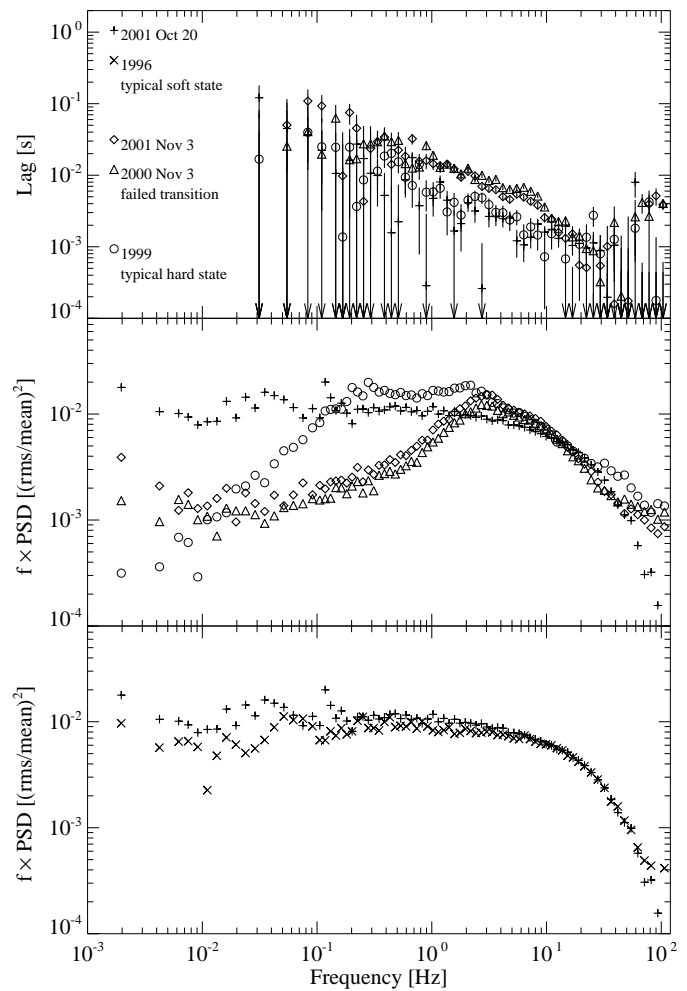}}
\caption{Time lags and PSD for the soft state of 2001~Oct
  compared with other typical soft state, hard state, and failed state
  transition data.}\label{fig:softcompare}
\end{figure}

During 2001~September, the RXTE-ASM rate rose from its typical
20--30\,cps to $\sim$100\,cps, with precursor flares in July and
August (Fig.~\ref{fig:octlc}). Reaching its peak in mid-September, the
soft X-ray flux first declined, and was found at a lower but still
enhanced ASM level of approximately 60\,cps in 2001~December, the end
of the time period considered here.  However, in contrast to previous
flares, this time a clear reduction of the radio emission was also
observed, with almost zero flux for about one week in the beginning of
October (Fig.~\ref{fig:octlc}).  After that time the radio emission
slowly turned on again, although continued flaring in both the radio
and the X-rays was observed\footnote{We note that the radio monitoring
  has been more frequent than our pointed X-ray observations, and this
  is the first clear indication of the soft state in both the radio
  and the X-rays since the radio monitoring began.}.  The combination
of the strong ASM flux and the missing radio emission is typical for a
soft state.  As shown in Fig.~\ref{fig:softcompare}, pointed RXTE
observations performed in 2001~October show a clear soft state
behavior: comparing PSDs and time lags for the two monitoring
observations of 2001 October~20 and 2001 November~3 (dashes in
Fig.~\ref{fig:octlc}) with typical data from failed state transitions
and from the 1996 soft state shows that on 2001 October~20 the PSD had
a clear $f^{-1}$ shape, turning over at $\sim$10\,Hz, an X-ray time
lag comparable with that measured during the standard hard state and
an X-ray coherence that was close to unity. The time lag behavior, the
unity coherence, and the PSD shape are typical for the soft state
\citep{cui:97b,pottschmidt:00a}.  The X-ray spectrum of the 2001
October~20 PCA observation in the 3--20\,keV range can be very well
described by the empirical soft state spectrum of \cite{cui:96a},
i.e., the sum of a black body and a broken power law. The broken power
law has a photon index of $\Gamma_1=2.78\pm0.03$ below, and
$\Gamma_2=2.10\pm 0.02$ above the break energy of
$E_{\text{b}}=11.1\pm 0.1$\,keV ($\chi^2=27.5$ for 31 degrees of
freedom, assuming an energy independent systematic uncertainty of
0.3\%, a reasonable value for the current calibration of the PCA). The
temperature of the black body is measured to $kT_{\text{BB}}=0.36\pm
0.03$\,keV. The black body contributes 20\% to the total 2--10\,keV
flux of $3.6\times 10^{-8}\,\text{erg}\,\text{cm}^{-2}\,\text{s}$.
These spectral parameters are similar to those of the 1996 soft state
\citep{cui:96a}, although the observed 2--10\,keV flux is a factor
$\sim$2 higher. We note that the 2001 October~20 data require an
additional and very broad iron line at 6.4\,keV (equivalent width
$\sim 440$\,eV with a Gaussian width $\sigma=0.9\pm 0.1$\,keV) that
had not been seen in 1996.

On 2001 October~22, RXTE performed a second observation close to this
soft state observation. Here, the PSD slightly deviated from the soft
state PSD, while the X-ray spectrum was still well described by the
empirical soft state model. Shortly after this, during the next
monitoring observation on 2001 November~3, the PSD showed an
``intermediate state'' shape similar to that during the failed state
transitions of 1999~November and 2000~December discussed above, i.e.,
with the $L_1$ component missing and an increased power law component.
Also pointing towards a transitional behavior are the increased time
lag and the reduced X-ray coherence.  Both the PSD and the X-ray time
lag are similar to the behavior measured on 2000 November~03, one year
before the 2001 soft state (see Fig.~\ref{fig:softcompare}). The X-ray
spectrum is much harder and can be well described by the hard state
spectral model.

We conclude that in 2001 October and November, Cyg~X-1 was switching
back and forth between the ``classic'' soft state and the
``intermediate state''. These changes from the soft state to the
transitional behavior happened very quickly, within a few days at
most. Furthermore, the very similar behavior of the source during the
failed state transitions and the 2001 October transition into the soft
state provides clear evidence that the ASM flares seen since 1996 are
indeed ``failed state transitions'', as opposed to being brief ``soft
states''. As we will show in a later paper \citep{wilms:03a}, after
that episode Cyg~X-1 settled into a soft state until the end of 2002.

\subsection{Narrow Lorentzians}\label{sec:thin}
In the previous sections we have described the overall evolution of
the four broad Lorentzians, which are responsible for most of the
variability observed in Cyg~X-1. Here we finish our
discussion by concentrating on the additional narrow Lorentzians,
$L_{\text{add}, 1}$ through $L_{\text{add}, 3}$, which were needed in
about 65\% of all observations to completely describe the observed
power spectra (Sect.~\ref{fits}). As these components are always rather
narrow, with $Q \gg 2$, we will follow the conventions outlined, e.g.,
by \citet{belloni:02a} and call these components QPOs.  As already
mentioned in Sect.~\ref{fits}, these QPOs fall into two
categories: rather narrow and clearly visible QPOs, $L_{\text{add},1}$
and $L_{\text{add},2}$, with frequencies above roughly 1\,Hz, and a
broader structure, $L_{\text{add},3}$, at frequencies below 0.2\,Hz.

\begin{figure*}

\includegraphics[width=12cm]{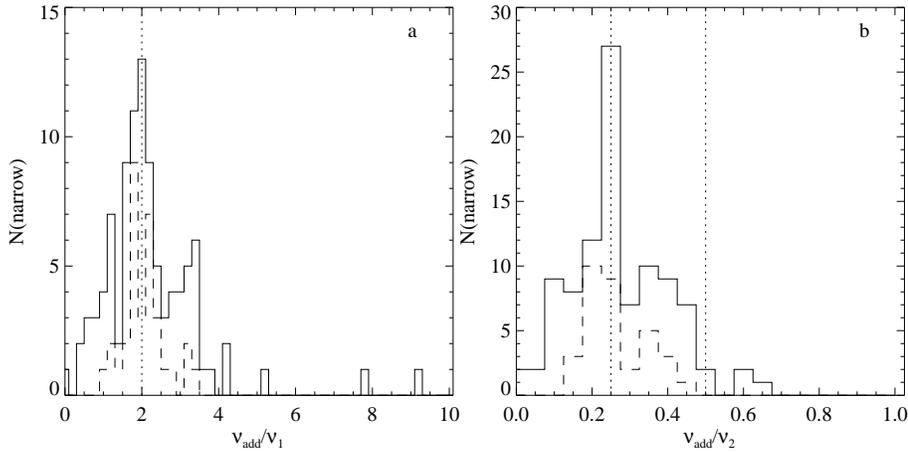}
\caption{\textbf{a} Number of detected thinner Lorentzian components as a
  function of $\nu_{\text{add}}/\nu_1$ and \textbf{b}
  $\nu_{\text{add}}/\nu_2$. The dashed lines represent the same
  distribution for QPOs with $Q\ge 50$ only. The dotted line in panel
  a corresponds to a frequency ratio of 2; in panel b, dotted lines
  designate frequency ratios of 1:2 and 1:4 with respect to
  $\nu_2$.\label{fig:thinfreqdist}}
\end{figure*}

Our emphasis on modeling the broad Lorentzians led us to
logarithmically rebin the measured PSDs, to the extent that a
determination of the parameters of the narrow features was not always
possible (see footnote~\ref{psdfoot}). For example, the logarithmic
rebinning of the PSDs means that the QPO was concentrated in only a
few frequency bins such that its $Q$-value could not be determined.
$Q$-factors larger than 50 in Table~\ref{tab:fitresults} generally are
an indication that these QPOs are concentrated in one frequency bin.
The majority of the remaining $Q$-factors cluster between $\sim$2 and
$\sim$15, i.e., these QPOs are somewhat broader structures.  Due to
these limitations in our modeling, we can not quote formal
uncertainties for the fit parameters of the QPOs.  We will
concentrate, therefore, in the following on the frequency behavior of
the QPOs, keeping in mind the above problems with the detection and
quantification of the QPO features. For consistency with the
description of the broad Lorentzians, we will continue to use the peak
frequency to characterize the frequency of the QPOs. Due to the large
$Q$-values involved, the peak frequency is usually very close to the
center frequency of the Lorentzian used to model the QPO.

Turning first to the two narrow QPOs, Fig.~\ref{fig:thinfreqdist}
shows the distribution of the ratio between the peak frequencies of
all narrow Lorentzians to those of $L_1$ and $L_2$. Note that the
majority of these additional PSD components are found between $L_1$
and $L_2$.  There are a few observations in which a QPO is present
below the peak frequency of $L_1$, while no additional components were
required above $L_2$. As is clearly shown in
Fig.~\ref{fig:thinfreqdist}a, an appreciable fraction of the QPOs are
found at the first harmonic of the \emph{peak} frequency of the $L_1$
component, $\nu_1$ ($\sim 20$\% of all detected QPOs have peak
frequencies in the range $1.8\le \nu_\text{add}/\nu_1 \le 2.2$). This
enhanced fraction of QPOs at $2\nu_1$ suggests a relationship with the
broad Lorentzian $L_1$.  Taken at face value, one might interpret the
QPOs in this frequency range as overtones of $L_1$, and indeed $L_1$
often shows a more complex structure which is modeled away by adding a
Lorentzian to the model. We note that such a presence of possibly
harmonically related components is not unexpected as it was seen
before by \citet{nowak:00a} in his treatment of the PSDs of Cyg~X-1
and GX~339$-$4.

We caution, however, that the ``overtone'' interpretation of this
substructure is not the only interpretation.  Since $L_1$ is broad,
one might expect its overtones to be broad as well, and this is not
the case: the narrow QPOs with $Q\ge 50$, i.e., those QPOs which are
only present in one frequency bin of our rebinned PSDs, have a
distribution which is similar to that of QPOs with $Q<50$
(Fig.~\ref{fig:thinfreqdist}a, dashed lines).  Furthermore, since for
the peak frequencies of $L_1$ and $L_2$, $\nu_2/\nu_1\sim 8$
(Table~\ref{tab:averages}), we have $2\nu_1 \sim \nu_2/4$, such that
these QPOs could also be subharmonics of $L_2$. We conclude,
therefore, that the frequency distribution of the QPOs is indicative
of a relationship between at least part of these components and the
broad Lorentzians, but that the quality of the current data is not
good enough to distinguish between the different interpretations.
This is especially true given that the Lorentzian shape assumed for
$L_1$ through $L_4$ might only be an approximation to the real shape
of the broad components that apparently comprise the PSD.

Finally, we turn to the broad additional Lorentzian present at
frequencies $<0.2$\,Hz, $L_{\text{add},3}$. In about half of the 17
observations in which this component was present, an additional power
law component was also required in the description of the data. Our
data indicate a very loose correlation between the $Q$-value of
$L_{\text{add},3}$ and its peak frequency.  As shown in the previous
sections, PSDs containing a power law component are one indicator of
(failed) state transitions. The presence of $L_{\text{add},3}$ in
these PSDs thus indicates that the PSD during the failed state
transitions is slightly more complicated than the simple sum of a
power law and the four broad Lorentzians and that the low frequency
PSD is more structured during these events (see, e.g.,
\citealt{miyamoto:94} and \citealt{belloni:96}).  Finally, we note
that we have not found statistically significant correlations between
the peak frequency of $L_{\text{add},3}$ and that of $L_1$ through
$L_4$ or the narrow QPOs . The quality of our PSDs at low frequencies
is not good enough, however, to allow a more detailed analysis.

\section{Discussion and Conclusion}\label{sec:disc}

\subsection{Summary}
The rapid high energy variability of galactic black holes is still not
understood.  The behavior of the PSD and the time lags during our
campaign, however, gives some clue as to its origin.

The most important result of this paper is that the shape of the PSD
during the hard state and during the intermediate state can be
productively interpreted with the decomposition of the PSD into the
four broad peaked components, which we model as Lorentzians, and one
power law. In contrast to earlier analyses, we were able to show that
the decomposition holds for a very wide range of source parameters,
even though the contributions of the power law and the Lorentzians are
very different in the different states.  \emph{Our result therefore
  adds further weight to the claim that the four broad Lorentzians are
  not just a convenient description of the PSD (merely more successful
  than the approximation of a broken power law), but that instead the
  broad components are the fundamental building blocks of the PSD of
  Cyg~X-1 and thus possibly also of the PSDs of other black hole
  candidates.} We emphasize that a statistically significant
discrimination between different shapes of the Lorentzians is not
possible with the available data (see Sects.\ \ref{fits}
and~\ref{sec:thin}, and \citealt{belloni:02a}), however, our study
shows that any alternative building blocks for the PSD must at least
be similar in shape to the broad components revealed by our analysis.

Furthermore, there are two fundamental results from our campaign on
the long term variability of the source. The first result is that the
change of the general long term behavior of Cyg~X-1 from a ``quiet
hard state'' to a ``flaring hard state'', with frequent intermittent
failed state transitions after 1998~May, coincided with a change in
the PSD shape and amplitude. After 1998~May, $L_3$ was clearly much
weaker than before.  At the same time, the whole PSD shifted towards
higher frequencies in a way that preserved the ratio between the peak
frequencies, a larger fraction of PSDs showed low frequency noise, the
X-ray lag increased, and the average X-ray spectrum softened.  The
tendency for a softening of BHC spectra to correlate with a frequency
shift of characteristic features of the PSD has also been seen in
other sources
\citep{dimatteo:99a,gilfanov:99a,revnivtsev:00a,kalemci:01a,nowak:01b},
and thus might be considered an intrinsic feature of \emph{all} BHCs.

The second result of our campaign concerns the change in the PSD shape
during the X-ray flares. We interpret these flares as transitions from
the hard state via the intermediate state into the soft state.  Most
of these transitions ``failed'', i.e., the transition stopped before
the soft state was reached. That the ``failed transitions'' are caused
by the same physical mechanism as the normal transitions into the soft
state was shown by the two instances during the campaign where a brief
soft state was observed (Sect.~\ref{sec:failed}).  In these cases,
before the soft state was reached and after it was left again, the
source behavior was indistinguishable from the behavior during the
``failed transitions''.

On the way into the intermediate state, $L_2$ and $L_3$ almost
remained constant, while $L_1$ -- and, to a lower significance, also
$L_4$ -- became remarkably weaker.  At the same time, the X-ray
spectrum softened and the whole PSD shifted to higher frequencies.
Furthermore, the X-ray time lag increases in the 3.2--10\,Hz frequency
band during the transitions and the coherence function decreases.
\emph{This is the frequency band where the PSD is dominated by the
  $L_2$ and, to a lesser extent, the $L_3$ components.} In those cases
where the soft state was reached, the X-ray lag spectrum and the
coherence function during the soft state were found to be very similar
to that of the normal hard state, while the  PSD, in contrast to
the typical hard state, showed its characteristic $f^{-1}$ shape with
a cut-off at $\sim$10\,Hz. During the soft state, no radio emission
was detected. Such a detailed study of the hard state and transitions
has to date only been possible for Cyg~X-1. We note, however, that
observations with a sparser coverage show similar behavior for
GX~339$-$4 \citep{nowak:01b} and for \object{XTE~J1650$-$50}
\citep{kalemci:02a}.

\begin{figure*}

\includegraphics[width=12cm]{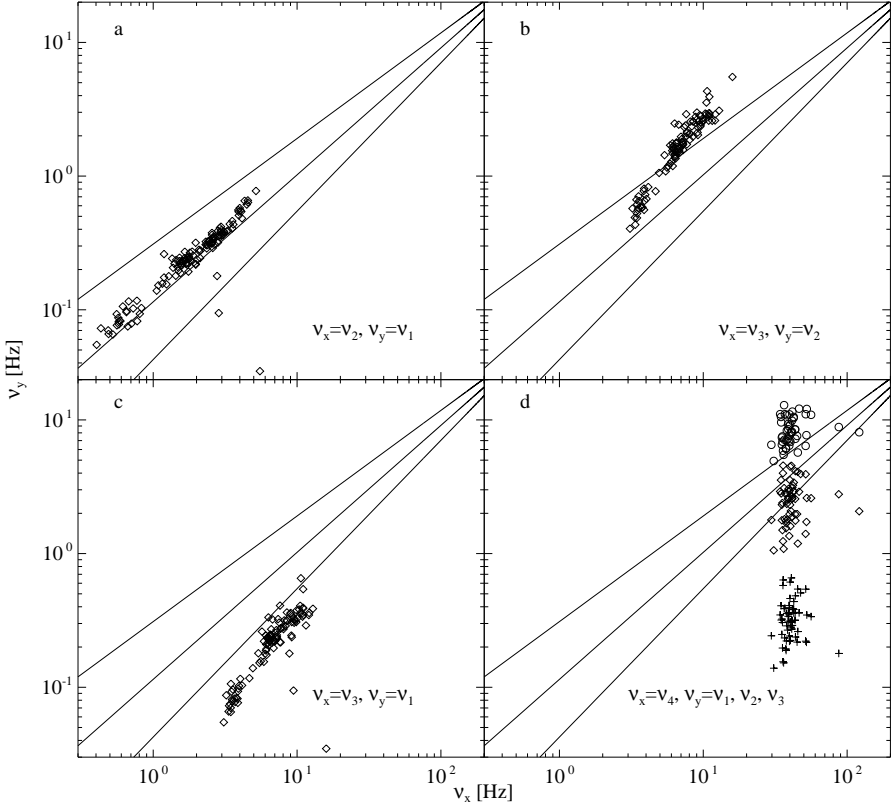}
\caption{\ Correlations between the peak frequencies of the broad noise
  components, $L_1$--$L_4$ and the frequency correlations discussed by
  \citet[][solid lines are
  $\nu_y=42\,\text{Hz}\,(\nu_x/500\,\text{Hz})^{0.95\pm
  0.16}$]{psaltis:99a}.
  \textbf{a)} $\nu_1$ versus $\nu_2$, 
  \textbf{b)} $\nu_2$ versus $\nu_3$, 
  \textbf{c)} $\nu_1$ versus $\nu_3$. 
  \textbf{d)} $\nu_1$ (crosses), $\nu_2$ (diamonds), and $\nu_3$
  (circles) versus the highest peak frequency, $\nu_4$. Only data
  taken in P40099 and P50110 are shown in panel~d to avoid biasing due to
  the 32\,Hz frequency cutoff in the P30157 data.}
\label{fig:cygfcorr}
\end{figure*}

\subsection{Relationship between Neutron Star and Black Hole Sources}
Apart from the analysis of \citet{nowak:00a}, Lorentzian modeling has
also been used for describing the power spectra of neutron star
sources. A recent summary of Lorentzian fits to data from neutron
stars and black holes has been recently published by
\citet{belloni:02a}. These authors base their work on their analysis
of RXTE observations of six neutron star sources
(\object{1E~1724$-$3045}, \object{SLX~1735$-$269},
\object{GS~1826$-$24}, \object{XTE~J1118$+$480}, \object{Cir X-1}, and
\object{GX~17$+$2}), as well as on published results from the black
holes Cyg~X-1, GX~339$-$4 \citep{nowak:01b}, and
\object{XTE~J1550$-$05} \citep{homan:01a}, and the neutron star
sources \object{4U~1915$-$05} \citep{boirin:00a} and
\object{4U~1728$-$34} \citep{disalvo:01b}. Furthermore,
\citet{vanstraaten:01a} have recently published an extensive analysis
of data from \object{4U~0614$+$09} and \object{4U~1728$-$34}.

\citet{belloni:02a} and \citet{vanstraaten:01a} find that one to six
Lorentzians are required to obtain a good description of the PSDs of
these sources. A slight complication when comparing these results to
ours is that \citet{vanstraaten:01a} and \citet{belloni:02a} mainly
use \emph{zero frequency centered} Lorentzians, i.e., shot-noise
profiles, and not Lorentzians with non-zero center frequencies.  As we
describe in Sect.~\ref{fits}, our choice appears to result in a
slightly lower $\chi^2$ \citep[but note the discussion in Sect.~2
of][]{belloni:02a}.  Despite the different fit functions, however, the
behavior of the maximum frequencies in both cases is still comparable.
In terms of ascending $\nu$, \citet{vanstraaten:01a} call $L_1$ and
$L_2$ the band-limited noise and zero-centered Lorentzian, $L_3$ the
low-frequency Lorentzian, and $L_4$ the hectohertz Lorentzian. In the
notation of \citet{belloni:02a}, $L_1$ corresponds to $L_{\text{b}}$
as this is the Lorentzian describing the break frequency of the PSD.
The Lorentzians $L_3$ and $L_4$, which describe the lower and upper
characteristic frequencies in the tail of the PSD, are called
$L_{\text{l}}$ and $L_{\text{u}}$. In many neutron star PSDs the $L_2$
component has a rather large $Q$, therefore, \citet{belloni:02a}
identify $L_2$ with the low frequency QPO, $L_{\text{LF}}$.

Note that the for all sources the ``correct'' identification of these
components is difficult.  For example, $L_1$ and $L_2$ are not present
together in all data sets such that one might misinterpret the band
limited noise (to stay in the terminology of
\citealt{vanstraaten:01a}) as the zero-centered Lorentzian.  Similar
problems also apply for Cyg~X-1: had the spacing of our observations
during the (failed) state transitions been coarser, we would not have
been able to correctly identify the Lorentzians during the state
transitions.

At higher frequencies, additional narrow Lorentzians are required to
describe the kilohertz Lorentzians in neutron star sources, however,
their connection to the PSDs of BHCs is less clear.  \citet{nowak:00a}
argue tentatively that $L_4$ could be identified with the upper
kilohertz QPO, however, the observed behavior of this component by
\citet{vanstraaten:01a} casts some doubt on this interpretation: while
neutron star kHz QPOs are varying in frequency, $L_4$ clearly does not
(e.g., Fig.~\ref{fig:norm}). On the other hand,
\citet{vanstraaten:01a} point out that there is a broad feature around
100\,Hz, the ``100\,Hz bump'', which is constant in frequency and
therefore a more probable counterpart to $L_4$.

There are two main correlations among frequency components that have
been claimed for neutron star and black hole sources (see
\citealt{belloni:02a} for a recent summary).  \citet{wijnands:99a}
have pointed out a correlation between the PSD break frequency and the
low frequency QPO of X-ray binaries.  These two frequencies might be
related to $\nu_1$ and $\nu_2$ in our classification scheme.
\citet{psaltis:99a} have pointed out a correlation between the low
frequency QPO and the lower frequency kHz QPO, i.e., frequencies that
could be identified with $\nu_2$ and $\nu_3$ in our scheme.  The
latter correlation especially relies on identifying these features in
multiple sources.  The origin of these correlations is unknown.

In Fig.~\ref{fig:cygfcorr} we plot various combinations of one
frequency versus another, and show that with the exception of $\nu_4$
\emph{all} frequency combinations show some sort of correlation (this
is just a consequence of the constancy of the frequency ratios shown
in Fig.~\ref{fig:ratio}).  Similar correlations for several sources
are also shown by \citet[][Figs.~11 and~12]{belloni:02a}, who point
out that -- while all different sources seem to show correlations
between individual Lorentzians -- these do not seem to completely
agree from source to source. Given the above two correlations, the one
that fits our data best is the correlation described by
\citet{psaltis:99a}, albeit applied to the $\nu_1$ versus $\nu_2$
correlation.  We note, however, that in this particular range of
frequencies, the $\nu_1$ versus $\nu_2$ correlation cited by
\citet{wijnands:99a} is numerically not very different than the
$\nu_2$-$\nu_3$ correlation of \citet{psaltis:99a}.  Likewise, the
$\nu_2$-$\nu_3$ data presented here are marginally consistent with the
correlation discussed by \citet{psaltis:99a}.  Although the individual
components discussed here span more than a decade in frequency, this
is not large enough a range to identify unambiguously given features
with the three decade wide correlations hypothesized and discussed by
\citet{psaltis:99a}.  Still, correlations among the frequency
components clearly do exist, and point to possible physical scenarios,
as we further discuss below.

\subsection{A Possible Physical Scenario}
We now discuss the empirical evidence from our campaign in terms of
more physical models for the accretion process in a black hole system.
The hard state X-ray spectrum is thought to be caused, at least in
part, by soft photons from the accretion disk which are Compton
upscattered in a hot electron gas \citep[][and references
therein]{sunyaev:79a,dove:97c,poutanen:99a}. In this picture, state
transitions are caused by the disappearance of this Comptonizing
medium. The cause for this disappearance is unknown, however, due to
the luminosity difference between the hard and the soft state, many
workers have assumed that the Compton corona is produced by some
physical process which can work only at mass accretion rates below a
certain threshold rate such as the magnetorotational instability
\citep{balbus:91}, advection \citep{esin:97a}, or some kind of coronal
outflow \citep[][and references therein]{merloni:01a,blandford:99a},
while others have claimed the presence of an up to now unknown
additional parameter which triggers the state transitions
\citep{homan:01a}. A similar picture of changing coronal properties
should also hold true within the hard state itself -- although the
Compton corona exists throughout the hard state, phases where a
different X-ray spectrum and PSD are observed might well correspond to
different coronal geometries
\citep[e.g.,]{nayakshin:99a,liu:99a,esin:97a,smith:01b}. Such a change
in the coronal configuration could explain the reduction of the $L_3$
component after 1998~May, as is also evidenced by an overall softening
of the spectrum since that time.

We note that for Cyg~X-1 the luminosity difference between the hard
and soft state is rather small \citep[$\sim$35\%;][]{zhang:97a}.
Furthermore, a power law component is always seen during the soft
state. This suggests that the soft state in Cyg~X-1 might not be a
``normal'' soft state as that seen in other black holes, but rather
closer to the very high state than the canonical soft state is.
Furthermore, there are cases where the soft state has not a higher
luminosity but a lower luminosity than the hard state. For
\object{GRS~1758$-$258}, an intermediate state has even been observed
at a higher luminosity than the hard state, while the soft state
appears to be at a much lower luminosity \citep{smith:01a}. A possible
explanation for these deviations from the canonical picture of the
soft state having a much higher luminosity than the hard state might
be an hysteresis effect \citep{miyamoto:95a}, however, we note that
regardless of the physical cause for deviations the observational
evidence still points towards changes in the accretion disk geometry.

That the accretion disk geometry changed around 1998~May might also be
the cause for the shift of the peak frequencies towards higher
frequencies observed during that time.  Even in prior analyses where
PSDs of BHC have been described in terms of power law fits and ``shot
noise models'' \citep[][and references therein]{lochner:91}, it has
been presumed that the characteristic frequencies can be attributed to
time scales of the accretion disk.  \citet{churazov:01a} have argued
that PSD ``breaks'' are related to the size scale at which the
accretion flow transits from a thin disk into a geometrically thick,
hot inner corona.  Di Matteo \& Psaltis (1999) \nocite{dimatteo:99a}
have argued that the dynamic range over which such a transition varies
is somewhat limited since the observed frequencies themselves,
although variable and correlated with spectral hardness, only span a
limited range.  Since flow frequencies scale as $\delta\nu/\nu \sim
(H/R)^2$ \citep[e.g.,][and references therein]{dimatteo:99a}, if the
Lorentzians are associated with resonant effects within the accretion
disk, they must originate from a region that is geometrically thicker
than the thin accretion disk, i.e., a comparably extended region such
as the accretion disk corona.

A simplified analysis of the characteristic frequencies of such a
transition region has been presented by \citet{psaltis:00a}.  These
authors note that if the outer disk acts as a ``noise source'', the
transition region acts as a ``filter'', and the inner corona acts as a
``response'' which processes the filtered noise into observable X-ray
variability, then one expects to observe a number of quasi-periodic
features.  These features would be related to radial, vertical,
acoustic, rotational, etc., oscillations of the transition region.
Many of these frequencies would be expected to scale as
$R_{\text{T}}^{-3/2}$, where $R_{\text{T}}$ is the transition region
radius.  More recently, \citet{nowak:01b} have shown that such a
scaling might exist for hard state observations of GX~339$-$4.
Specifically, they show that in ``sphere and disk'' coronal models, the
``coronal compactness'' -- where higher compactness,
$\ell_{\text{c}}$, means harder spectra -- scales as $\ell_{\text{c}}
\propto R_{\text{T}}$. Furthermore, the characteristic PSD
frequencies, measured in a manner similar to that used to describe the
Cyg~X-1 PSDs presented here, scale as $\ell_{\text{c}}^{-3/2}$.  Both
Cyg~X-1 and GX~339$-$4, in their hard states, show that spectral
softening is correlated with higher PSD frequencies, although we have
yet to characterize the Cyg~X-1 spectra in terms of the coronal models
presented by \citet{nowak:01b}.  Such an analysis is currently
underway \citep{gleiss:02a}.

Further hints towards attributing at least some of the Lorentzians to
the Compton corona come from the PSD behavior during the transitions
into the intermediate or the soft state. We hypothesize that the
variability causing $L_1$ and possible $L_4$ is also associated with
the accretion disk corona, since these components -- like the corona
-- appear to vanish during the transition.  Since $L_2$ and $L_3$
remain present in the 3.2--10\,Hz band, we are tempted to attribute the
increased X-ray time lag in this frequency band to these components.  As
discussed by \citet{nowak:00a}, the overall measured time lag may in
reality be a composite of time lags inherent to each individual PSD
component.  Whereas the composite time lag might be fairly small, the
time lags associated with the individual PSD components might be
somewhat larger \citep[specifically, see Fig.~5 of][]{nowak:00a}.  The
fairly long intrinsic time lags of these individual PSD components
would only be revealed in the absence of the other PSD components.
Such a picture might explain the larger X-ray time lag during state
transitions: Here, the $L_2$ and $L_3$ components, which sit in the
3.2--10\,Hz band, contribute a larger \emph{relative} fraction of the
total source variability. If their intrinsic lag is larger, a larger
lag would be observable during state transitions when the contribution
of $L_1$ and $L_4$ is reduced.  We note that such a picture of a
``composite X-ray time lag'' is somewhat suggested by the
observational shape of the time lag spectrum, where ``jumps'' between
different frequency bands of roughly constant time lag seem to
correspond to the overlap regions of the broad Lorentzians
\citep{nowak:98c,poutanen:01a,pottschmidt:00a}.

While the origin of the time lag is not yet understood, it is possible
that the time lag is somehow related to a characteristic length scale
of the medium which produces the observed photons. If this is true,
then the larger lags during state transitions may be showing that some
part of the source geometry changes and becomes larger during the
transitions.  It is important to note that although there is no radio
emission during the soft state, there does seem to be a slight
increase of the radio flux at least \emph{during} some of successful
transitions into and out of the soft state \citep{corbel:00a,zhang:97c}.
Furthermore, intermittent, rapid radio flaring activity also seems to
be associated with the (failed) transitions (Pooley, priv. comm.).

In general, we see that the time lags are correlated with the PSD
frequencies and correlated with the photon index, $\Gamma$
(Fig.~\ref{fig:multicorr}a in Sect.~\ref{sec:evolution}). 
For GX~339$-$4, spectral fits with the coronal model of \citet{coppi:99a}
-- where a spherical corona with seed photons distributed throughout it
according to the diffusion equation is modeled -- suggest an
\emph{increasing} radius with a softening spectrum \citep{nowak:01b},
consistent with a geometric interpretation of increased time lags.
\citet{nowak:01b} therefore argued for a jet-like model for the hard state
of GX~339$-$4 where the radius of the \emph{base} of the jet followed the
radius-frequency correlations suggested by the ``sphere and disk'' coronal
model (i.e., smaller \emph{base} radius yielding softer spectra), while the
\emph{vertical} extent of the corona/jet followed the correlations
suggested by the Coppi coronal model (i.e., more vertically extended
coronae yielding softer spectra).  Thus, frequencies might be generated at
the base of the jet, while time lags could be generated via propagation
along the jet.  Given the similar, but much more detailed, correlations
observed here in Cyg~X-1 it is tempting to ascribe the same type of model
to these data.  Whether the analogy to GX~339$-$4 holds in detail will
depend upon the results of detailed coronal model fits \citep{gleiss:02a}.

Along these lines, based mainly on observations of microquasars,
\citet{fender:01a} and \citet{fender:99b} have recently attempted to unify
the radio and X-ray observations into a geometrical picture of the region
surrounding the central black hole. We have previously suggested that the
enhanced X-ray lags during the ``flares'' observed in Cyg~X-1 seem to add
credibility to this picture \citep{pottschmidt:00b}. During the normal hard
state, there is a weak outflow of material from the region surrounding the
black hole. This outflow is responsible for producing the observed
synchrotron radio emission. We think it likely that the base of this
outflow coincides with the accretion disk ``corona'' that produces the hard
state X-ray spectrum and is, given the results above, also responsible for
part of the observed X-ray variability.  During the ``flares'', part of the
corona gets ejected. Thus the mass of the radio outflow is temporarily
enhanced, increasing the radio luminosity. As the corona is disturbed,
whatever resonant mechanism producing the $L_1$ and possibly also the $L_4$
variability will be disturbed, resulting in a decrease of its contribution
to the total rms variability.  Furthermore, if this ``corona'' is
``stretched'' because of the ejection, it seems likely that a temporary
increase in the X-ray lags should be associated with the outflow as the
characteristic length scales increase.  During the soft state, the corona
has fully vanished such that no radio outflow is observed and the X-ray
spectrum is soft.  If $L_2$ and $L_3$ are produced at the corona-disk
interface, they also vanish.

This very rough picture leaves many detailed questions open, such as
the explanation for the strong domination of the $f^{-1}$ power law
during the soft state, its cutoff frequency, the generation of the
intrinsic lag of the Lorentzians, or the physical mechanisms
responsible for the PSD itself.  However, we believe that monitoring
campaigns such as this one, that are able to assess the long term,
systematic changes in accreting systems, will continue to constrain
the parameters of the observed physical systems and will enable us to
switch from describing the ``accretion disk weather'' towards
understanding the ``accretion disk climate''.  Equally importantly,
X-ray astronomy is entering a phase of high spectral resolution
observations with such instruments as \textsl{Chandra} and
\textsl{XMM-Newton}.  The nature of these instruments does not allow
detailed and intensive monitoring campaigns such as this one.
However, given the continued operation of the \textsl{RXTE-ASM} and
judicious use of simultaneous pointed \textsl{RXTE} observations,
campaigns such as this will allow these individual high resolution
observations to be placed within their proper `global' context.

\begin{acknowledgements}
  We thank Emrah Kalemci for his contributions to the Fourier analysis
  tools, Craig Markwardt for his $\chi^2$ minimization routines, and
  Sara Benlloch, Tomaso Belloni, Rob Fender, Dimitrios Psaltis, Steve
  van Straaten, Michiel van der Klis, and Juri Poutanen for helpful
  discussions and comments on the manuscript. We are indebted to the
  RXTE schedulers, most notably Evan Smith, for making such a long
  multi-wavelength campaign feasible. This work has been partly
  financed by grants Sta~173/25-1 and Sta~173/25-3 of the Deutsche
  Forschungsgemeinschaft, by NASA grants NAG5-3072, NAG5-3225,
  NAG5-7265, NAS5-30720, National Science Foundation travel grant
  INT-9815741, and by a travel grant from the Deutscher Akademischer
  Austauschdienst.
\end{acknowledgements}

\bibliographystyle{aa}

\end{document}